\documentclass[prd,preprint,superscriptaddress,showpacs]{revtex4}
\usepackage{revsymb}
\usepackage[dvips]{graphicx}

\usepackage{epsfig}
\usepackage{graphicx}
\usepackage{indentfirst}
\usepackage{amsmath}
\usepackage{amsfonts}
\usepackage{amssymb}

\begin{document}
\begin{titlepage}
\title{Scalar models for the  generalized Chaplygin gas and the structure formation constraints}
\author{J\'ulio C. Fabris}
\email{E-mail: fabris@pq.cnpq.br}
\author{Thaisa C.C. Guio}
\email{E-mail:thaisa_guio@yahoo.com.br}
\author{Mahamadou Hamani Daouda}
\email{E-mail: daoudah8@yahoo.fr}
\author{Oliver F. Piattella}
\email{E-mail: oliver.piattella@gmail.com}
\affiliation{Universidade Federal do Esp\'{\i}rito Santo,
Departamento de F\'{\i}sica\\
Av. Fernando Ferrari, 514, Campus de Goiabeiras, CEP 29075-910,
Vit\'oria, Esp\'{\i}rito Santo, Brazil}

\date{\today}

\begin{abstract}
The generalized Chaplygin gas model represents an attempt to unify dark matter and dark energy. It is characterized by
a fluid with an equation of state $p = - A/\rho^\alpha$. It can be obtained from a generalization of the DBI action for
a scalar, tachyonic field. At background level, this model gives very good results, but it suffers from many drawbacks at
perturbative level. We show that, while for background analysis it is possible to consider any value for $\alpha$, the
perturbative analysis must be restricted to positive values of $\alpha$. This restriction can be circumvented if the
origin of the generalized Chaplygin gas is traced back to a self-interacting scalar field, instead of the DBI action.
But, in doing so, the predictions coming from formation of large scale structures reduce the generalized Chaplygin gas
model to a kind of quintessence model, and the unification scenario is lost, if the scalar field is the canonical one. However, if the unification condition is
imposed from the beginning as a prior, the model may remain competitive. More interesting results, concerning the unification program, are
obtained if a non-canonical self-interacting scalar field, inspired by Rastall's theory of gravity, is imposed. In this case, an agreement with the background
tests is possible.
\end{abstract}
\maketitle
\end{titlepage}
\section{Introduction}

A lot of effort has been devoted in the last decades to determine the matter-energy content of the universe. This effort concentrates not only
on the determination of the amount of the different components of the cosmic budget, but also on identifying their nature. The different components predicted by the
standard model of elementary particles (baryons, radiation and neutrinos) are believed to be known with high precision. Expressed in
terms of the ratio to the density necessary to have a flat spatial section (the critical density), the recent estimations lead to a baryonic parameter density today given by
$\Omega_{b0} = 0.0456\pm 0.0016$,
while the radiation density is given by $\Omega_{\gamma0} \sim 5\times10^{-5}$ \cite{komatsu}. The neutrino density depends on the number
of neutrino species and their masses, but estimations lead to values close to that found for radiation. Summing up all these contributions,
the so-called ordinary matter (including under this concept radiation and neutrinos) give a very low mass compared with the critical mass.
However, the general features of the spectrum of anisotropies in cosmic microwave background radiation imply that the total matter
density of the universe must be very close to the critical one. In the reference \cite{komatsu}, using 7-years WMAP measurements, Baryonic Acoustic Oscillations (BA0) and Supernova type Ia (SN Ia) data, the curvature parameter is found to be
$\Omega_{k0} = -0.0057^{+0.0067}_{-0.0068}$: the spatial section of the universe is essentially flat, and the
total density of mass/energy existing in the universe must be very close to the critical value. Hence, most of the matter/energy of the universe comes from the contributions not predicted
by the standard model of elementary particles.
\par
The dynamics of galaxies and clusters of galaxies presents important anomalies, requiring the presence of a pressureless, non-baryonic component,
dubbed dark matter, to be explained. The anisotropy spectrum of CMB and the present stage of accelerated universe require the presence of
another fluid, in principle exhibiting negative pressure, dubbed dark energy. For a recent review of these evidences, see reference \cite{caldwell}.
The main quoted candidates to be the constituents of dark matter are axions and neutralinos \cite{bertone}, but these candidates come from fundamental
physical theories not yet tested experimentally. For dark energy, the most natural candidate is the cosmological constant \cite{padma}. An important
alternative to the cosmological constant as the dark energy component is a dynamical scalar field called quintessence \cite{martin}. All these
proposals to the dark sector of the universe have their advantages and their drawbacks, that are exposed in the quoted references.
\par
A quite different alternative to describe the dark sector of the universe is the so-called unified scenario where both components are represented by a unique component playing at same time the r\^oles of dark matter and dark energy. The prototype of the unification scenario is the Chaplygin gas (CG) model
\cite{moschella,berto1,neven,fabris,jackiw} which, in its generalized form, reads
\begin{equation}
\label{eos}
p = - \frac{A}{\rho^\alpha},
\end{equation}
where $\alpha$ and $A$ are two free constants.
When $\alpha = 1$ we recover the original Chaplygin gas model, which has an interesting connection with the Nambu-Goto action \cite{jackiw}.
Using the Friedmann-Lema\^{\i}tre-Robertson-Walker
metric and inserting the equation of state (\ref{eos}) into the conservation law for a fluid 
\begin{equation}
\dot\rho + 3\frac{\dot a}{a}(\rho + p) = 0,
\end{equation}
(dot means derivative with respect to the cosmic time)
we can express $\rho$ as function of the scale factor $a$:
\begin{equation}
\rho = \biggr[A + \frac{B}{a^{3(1 + \alpha)}}\biggl]^\frac{1}{1 + \alpha},
\end{equation}
where $B$ is an integration constant,
indicating that the GCG behaves as pressureless matter for $a \rightarrow 0$ and as a cosmological constant when $a \rightarrow \infty$.
This is the main idea of the unification program.
\par
The Generalized Chaplygin Gas (GCG) model has been tested against many different observational data, like type Ia Supernova(SN Ia) \cite{colistete}, Cosmic Microwave Background Radiation (CMB) \cite{berto2,finelli,piattella1}, matter power spectrum (PS) \cite{piattella2,hermano}. The general scenario emerging from these
different tests is not uniform, depending on the priors and on the statistics scheme. Initially, it has been argued that the CMB tests \cite{berto2,finelli} favor a scenario around $\alpha = 0$, which essentially reduces the GCG model to the $\Lambda$CDM model. But, 
in reference \cite{piattella1}, by using the integrated Sachs-Wolfe effect, models with $\alpha \sim 300$ seem also to be favored.
On the other hand, SN Ia tests favor negative values of $\alpha$ \cite{colistete}. The perturbative analysis for negative values of $\alpha$
is not allowed due to resulting negative values of the square adiabatic sound velocity \cite{hermano}
\begin{equation}
c_a^2 = \frac{\dot p}{\dot\rho} = \frac{\alpha A}{\rho^{\alpha + 1}}.
\end{equation}
\par
We intend to address in the present work the last feature of the GCG model. Is it possible to consider negative values of $\alpha$ using a perturbative
analysis? Negative square sound velocity appears, for $\alpha < 0$, in the fluid representation of the GCG model.
The original, more fundamental, representation of the GCG model, is given by DBI action \cite{jackiw,berto1}, which
employs a scalar, tachyonic field. In reference \cite{eduardo}
it has been shown that, for $\alpha = 1$ (original Chaplygin gas model) the DBI action gives the same perturbative expressions as those found in the
fluid representation, but that such equivalence, at perturbative level, is broken if a self-interacting scalar field is used to
represent the CG model instead of the DBI action.
\par
In the present work, we extend the analysis to the case of the GCG model. It will be shown that the complete equivalence of the DBI action
and the fluid representation remains for $\alpha \neq 1$. A canonical self-interacting scalar model for the GCG model will be developed, and in
this case, the equivalence is lost at perturbative level. Hence the perturbative analysis can be made even for $\alpha < 0$. The comparison with the observational data will show, however,
that positive values of $\alpha$ will be favored, except perhaps when the unification scenario is imposed from the beginning. All analysis will be made at perturbative level,
since the background is the same as in the fluid or DBI representations.
\par
The canonical self-interacting scalar field, however, hardly can represent a unification model for dark matter and dark energy, due to one
crucial property: its sound velocity is equal to one (in units of velocity of light). Any model for dark matter must have a zero effective speed of sound
by the end of the radiative era. However, a non-canonical self-interacting scalar field, inspired by the Rastall's theory of gravity \cite{rastall}, may change
the overall picture. This non-canonical scalar field obeys a dynamical equation very different from the Klein-Gordon one, and at same time it can
not be reduced to a kind of $K$-essence model, and its effective sound velocity can be set equal to zero \cite{liddle}. The Rastall's theory is based essentially on a modification of the conservation law of the energy-momentum tensor. When, a specific representation for this scalar field is chosen, reproducing the background results for the GCG model,
the unification scenario leads to compatible results even at perturbative level. This is the main result of this paper.
\par
In the next section, we show the equivalence of the fluid representation with the DBI representation for the GCG model at the background level.
The equivalence is extend to the perturbative level in section III. An equivalent canonical scalar model is developed in section IV, and the perturbative
analysis is performed in section V. This analysis is reproduced for the Rastall's model for the scalar field in section VI, showing that
a coherent scenario can emerge in this case. In section VII we present our conclusions.

\section{The DBI formulation}

The DBI Lagrangian is given by the following expression:
\begin{equation}
\label{lagr}
{\cal L} = V(T)\sqrt{1 - T_{;\rho}T^{;\rho}},
\end{equation}
where $T$ is a scalar (tachyonic) field, and $V(T)$ is the potential for this field. An energy-momentum tensor can be constructed from this
action.
Variation with respect to the metric leads to the following expression for the energy-momentum tensor:
\begin{equation}
T_{\mu\nu} =  - \frac{2}{\sqrt{-g}}\frac{\delta S}{\delta g^{\mu\nu}} = \frac{V(T)\partial_\mu T\partial_\nu T}{\sqrt{1 - T_{;\rho}T^{;\rho}}} + V(T)\sqrt{1 - T_{;\rho}T^{;\rho}}g_{\mu\nu},
\end{equation}
where $S$ is the action constructed from the Lagrangian (\ref{lagr}).
Comparing with the energy-momentum tensor of a perfect fluid,
\begin{equation}
\label{pf}
T_{\mu\nu} = (\rho + p)u_\mu u_\nu - p g_{\mu\nu},
\end{equation}
and using the homogeneous, isotropic FLRW metric
\begin{equation}
ds^2 = dt^2 - a(t)^2\gamma_{ij}dx^i dx^j,
\end{equation}
we obtain the following expressions for the density and pressure:
\begin{equation}
\rho_T = \frac{V(T)}{\sqrt{1 - \dot T^2}}, \quad p_T = - V(T)\sqrt{1 - \dot T^2}.
\end{equation}
This leads to the equation of state
\begin{equation}
\label{eos-chap}
p_T = - \frac{V(T)^2}{\rho_T}.
\end{equation}
For $V(T) = \sqrt{A} = \mbox{constant}$,
the equation of state (\ref{eos-chap}) represents the traditional Chaplygin gas model. When $V(T)$ is not a constant new possibilities are opened,
and the richness of the model has been pointed out in reference \cite{gorini}.
\par
The GCG, characterized by the equation of state
\begin{equation}
\label{eos-gchap}
p = - \frac{A}{\rho^\alpha},
\end{equation}
can be obtained from the action
\begin{equation}
\label{gta}
{\cal L}_T = V(T)\biggr[1 - (\partial_\rho T\partial^\rho T)^\frac{1 + \alpha}{2\alpha}\biggl]^\frac{\alpha}{1 + \alpha},
\end{equation}
where $V(T) = A^\frac{1}{1 + \alpha} =$ constant.
In fact, from (\ref{gta}), we have the energy-momentum tensor
\begin{eqnarray}
T_{\mu\nu} = \frac{V(T)(\partial_\rho T\partial^\rho T)^\frac{1 - \alpha}{2\alpha}\partial_\mu T\partial_\nu T}{\biggr[1 - (\partial_\rho T\partial^\rho T)^\frac{1 + \alpha}{2\alpha}\biggl]^\frac{1}{1 + \alpha}} + V(T)\biggr[1 - (\partial_\rho T\partial^\rho T)^\frac{1 + \alpha}{2\alpha}\biggl]^\frac{\alpha}{1 + \alpha}g_{\mu\nu}.
\end{eqnarray}
A direct comparison with the perfect fluid energy-momentum tensor (\ref{pf}), leads to
\begin{eqnarray}
\label{rho-dbi}
\rho_T &=& \frac{V(T)}{\biggr[1 - (\partial_\rho T\partial^\rho T)^\frac{1 + \alpha}{2\alpha}\biggl]^\frac{1}{1 + \alpha}},\\
\label{p-dbi}
p_T &=& - V(T)\biggr[1 - (\partial_\rho T\partial^\rho T)^\frac{1 + \alpha}{2\alpha}\biggl]^\frac{\alpha}{1 + \alpha},\\
\label{v-dbi}
u_\mu &=& \frac{\partial_\mu T}{\sqrt{\partial_\rho T\partial^\rho T}}.
\end{eqnarray}

\section{Perturbations of the generalized DBI action}

The perturbation of the perfect fluid energy-momentum tensor (\ref{pf}) is now carried out using the synchronous coordinate condition
$h_{\mu0} = 0$. The components of the perturbed energy-momentum tensor are
\begin{eqnarray}
\label{fa}
\delta T_{00} &=& \delta\rho,\\
\label{fb}
\delta T_{0i} &=& (\rho + p)\delta u_i,\\
\label{fc}
\delta T_{ij} &=& - p h_{ij} - \delta p g_{ij}.
\end{eqnarray}
The perturbation of the generalized DBI action (\ref{gta}) leads, on the other hand, to the following expressions when $V(T) = V =$ constant:
\begin{eqnarray}
\label{pa-dbi}
\delta T_{00} &=& \frac{1}{\alpha}\frac{V\dot T^\frac{1}{\alpha}\delta\dot T}{\biggr[1 - \dot T^\frac{1 + \alpha}{\alpha}\biggl]^\frac{2 + \alpha}{1 + \alpha}},\\
\label{pb-dbi}
\delta T_{0i} &=& \frac{V\dot T^\frac{1}{\alpha}\partial_i\delta T}{\biggr[1 - \dot T^\frac{1 + \alpha}{\alpha}\biggl]^\frac{1}{1 + \alpha}},\\
\label{pc-dbi}
\delta T_{ij} &=& - \frac{V\dot T^\frac{1}{\alpha}\delta\dot T}{\biggr[1 - \dot T^\frac{1 + \alpha}{\alpha}\biggl]^\frac{1}{1 + \alpha}}g_{ij}
+ V(T)\biggr[1 - \dot T^\frac{1 + \alpha}{\alpha}\biggl]^\frac{\alpha}{1 + \alpha}h_{ij}.
\end{eqnarray}
\par
From the general form of the density, pressure and velocity obtained from the energy-momentum tensor of the DBI action, (\ref{rho-dbi},\ref{p-dbi},\ref{v-dbi}), we obtain, in the synchronous coordinate condition,
\begin{eqnarray}
\label{drho-dbi}
\delta\rho_T &=& \frac{1}{\alpha}\frac{V(T)\dot T^\frac{1}{\alpha}\delta\dot T}{\biggr[1 - \dot T^\frac{1 + \alpha}{\alpha}\biggl]^\frac{2 + \alpha}{1 + \alpha}},\\
\label{dp-dbi}
\delta p_T &=& \frac{V(T)\dot T^\frac{1}{\alpha}\delta\dot T}{\biggr[1 - \dot T^\frac{1 + \alpha}{\alpha}\biggl]^\frac{1}{1 + \alpha}},\\
\label{dv-dbi}
\delta u_i &=& \frac{\delta T_{,i}}{\dot T}.
\end{eqnarray}
Inserting the expressions (\ref{drho-dbi},\ref{dp-dbi},\ref{dv-dbi}) into (\ref{pa-dbi},\ref{pb-dbi},\ref{pc-dbi}), we obtain
(\ref{fa},\ref{fb},\ref{fc}). Hence, the perturbation of the generalized DBI energy-momentum tensor is perfectly equivalent to its fluid counterpart.
This equivalence has been verified for the first time (to our knowledge) in reference \cite{frolov}.
It is direct to verify that,
\begin{equation}
\frac{\delta p_T}{\delta\rho_T} = \frac{\dot p_T}{\dot\rho_T},
\end{equation}
as far as the potential is constant. The case were $V(T)$ is not constant has been analyzed, for some specific forms of the potential, in
reference \cite{frolov}.
In this sense, $\alpha$ negative implies a negative squared sound velocity. This can be verified taking the ratio between (\ref{dp-dbi})
and (\ref{drho-dbi}). In this sense, the extension of the perturbative analysis for negative values of $\alpha$ is forbidden, in
opposition to what happens with the background tests, like SNIa or BAO, where the problems connected with the sound velocity do not appear.

\section{Scalar model}

In order to exploit the possibility of $\alpha < 0$ we must abandon the DBI framework. The most general framework besides the DBI is
the self-interacting scalar field. If this self-interacting scalar field is implemented in the canonical way, there is the mentioned
problem with the sound speed: $c_s^2 = 1$. Moreover, while the DBI action is complete equivalent to the Chaplygin gas model, the
self-interacting scalar field presents many challenges, as non-unicity and stability of trajectories, requiring specific initial
conditions \cite{gori}. For the moment, we ignore all these problems and consider the canonical self-interacting scalar field
as an effective model which asks for a fundamental description before the decoupling between matter and radiation, as well as particular initial
conditions.
This possibility has been touched in \cite{eduardo}, but fixing $\alpha = 1$. But, at least for this
case, it has been found that neither the unification scenario \cite{colistete} nor the anti-unification scenario \cite{hermano} have been resulted from this self-interacting
scalar field approach: the matter density parameter was in fact very close to that predicted by the $\Lambda$CDM or the quintessential model.
The aim of the present investigation is to verify the predictions when all possible values of $\alpha$ are considered. 
\par
The action is now given by
\begin{equation}
L = \frac{1}{16\pi G}\sqrt{- g}\biggr[R - \phi_{;\rho}\phi^{;\rho} + 2V(\phi)\biggl] + L_m,
\end{equation}
where $V(\phi)$ defines the potential and $L_m$ is the matter Lagrangian.
The field equations are
\begin{eqnarray}
R_{\mu\nu} - \frac{1}{2}g_{\mu\nu}R &=& 8\pi GT_{\mu\nu} + \phi_{;\mu}\phi_{;\nu} - \frac{1}{2}g_{\mu\nu}\phi_{;\rho}\phi^{;\rho} + g_{\mu\nu}V(\phi),\\
{T^{\mu\nu}}_{;\mu} &=& 0,\\
\square\phi &=& - V_\phi,
\end{eqnarray}
where the subscript $\phi$ indicates derivative with respect to the scalar field.
\par
The equations of motion are
\begin{eqnarray}
3\biggr(\frac{\dot a}{a}\biggl)^2 &=& 8\pi G\rho_\phi = \frac{\dot\phi^2}{2} + V(\phi),\\
2\frac{\ddot a}{a} + \biggr(\frac{\dot a}{a}\biggl)^2 &=& - 8\pi G p_\phi = - \frac{\dot\phi^2}{2} + V(\phi).
\end{eqnarray}
In the generalized Chaplygin gas model, the energy density behaves as
\begin{equation}
\label{def}
\rho_c = \rho_{c0}g(a)^\frac{1}{1 + \alpha}, \quad g(a) = \bar A + \frac{(1 - \bar A)}{a^{3(1 + \alpha)}}.
\end{equation}
This expression may be rewritten as
\begin{equation}
\Omega_c = \Omega_{c0}g(a)^\frac{1}{1 + \alpha}, \quad \Omega_c = \frac{8\pi G\rho_c}{3H_0^2}, \quad \Omega_{c0} = \frac{8\pi G\rho_{c0}}{3H_0^2}.
\end{equation}
Making the substitution
\begin{equation}
tH_0^2 \rightarrow t, \quad \frac{V(\phi)}{H_0^2} \rightarrow V(\phi),
\end{equation}
we may write
\begin{eqnarray}
\rho_\phi &=& \frac{\dot\phi^2}{2} + V(\phi) = 3\Omega_{c0}g^\frac{1}{1 + \alpha},\\
p_\phi &=& \frac{\dot\phi^2}{2} - V(\phi) = - 3\Omega_{c0}\bar Ag^{-\frac{\alpha}{1 + \alpha}}.
\end{eqnarray}
Hence, we obtain
\begin{eqnarray}
\dot\phi &=&  \sqrt{3\Omega_{c0}}\biggr\{g^\frac{1}{1 + \alpha} - \bar Ag^{-\frac{\alpha}{1 + \alpha}}\biggl\}^{1/2},\\
V(\phi) &=&  \frac{3\Omega_{c0}}{2}\biggr\{g^\frac{1}{1 + \alpha} + \bar Ag^{-\frac{\alpha}{1 + \alpha}}\biggl\}.
\end{eqnarray}
This scalar field model reproduces exactly the background of the generalized Chaplygin gas in presence of baryonic matter. When baryons are 
absent, we find the potential of reference \cite{berto1}.

\section{Perturbative analysis of the scalar model}

In order to perform a perturbative study of this model, we rewrite the Einstein's equations in the following form:
\begin{eqnarray}
R_{\mu\nu} &=& 8\pi G\biggr(T_{\mu\nu} - \frac{1}{2}g_{\mu\nu}T\biggl) + \phi_{;\mu}\phi_{;\nu} - g_{\mu\nu}V(\phi),\\
\square\phi &=& - V_\phi(\phi),\\
{T^{\mu\nu}}_{;\mu} &=& 0.
\end{eqnarray}
Again, we choose to employ the synchronous coordinate condition $h_{\mu0} = 0$ in carrying out the perturbative study. Since all relevant
modes are well inside the cosmological horizon the final results do not depend on this choice.
\par
The standard perturbative calculation using the synchronous coordinate condition leads to the final set of equations:
\begin{eqnarray}
\label{pea}
\ddot\delta + 2\frac{\dot a}{a}\dot\delta - \frac{3}{2}\Omega_m\delta &=& 2\dot\phi\dot\lambda - 2V_\phi\lambda,\\
\label{peb}
\ddot\lambda + 3\frac{\dot a}{a}\dot\lambda + \biggr[\frac{k^2}{a^2} + V_{\phi\phi}\biggl]\lambda &=& \dot\phi\dot\delta.
\end{eqnarray} 
In these equations, $\delta = \delta\rho/\rho$ is the density contrast, $\lambda = \delta\phi$, the parameter $k$ is the wavenumber resulting
from the Fourier decomposition of the perturbed quantities, and $\Omega_m = \Omega_{m0}/a^3$ is the density parameter for the matter
component, $\Omega_{m0}$ being its value today. This density parameter contains two terms, the baryonic component and the dark matter
component $\Omega_{m0} = \Omega_{b0} + \Omega_{dm0}$. Equations (\ref{pea},\ref{peb}) refer
to the Fourier transform of the fundamental quantities, that is, we should more properly write $\delta_k$ and $\lambda_k$.
\par
In order to carry out the comparison with observations, it is more convenient to re-express equations (\ref{pea},\ref{peb}) using the
scale factor as dynamical variable. Moreover, we divide both equations by $H_0^2$, the Hubble parameter today. The final (dimensionless) equations
are the following:
\begin{eqnarray}
\delta'' + \biggr(\frac{2}{a} + \frac{f'}{f}\biggl)\delta' - \frac{3}{2}\frac{\Omega_m}{f^2}\delta &=& 2\frac{\dot\phi}{f}\lambda' - \frac{V_\phi}{f^2}\lambda,\\
\lambda'' + \biggr(\frac{3}{a} + \frac{f'}{f}\biggl)\lambda' + \biggr[\biggl(\frac{kl_0}{af}\biggl)^2 + \frac{V_{\phi\phi}}{f^2}\biggl]\lambda
&=& \frac{\dot\phi}{f}\delta',
\end{eqnarray}
where the primes mean derivative with respect to $a$ and $l_0 = 3.000 h^{-1}\,Mpc$ is the Hubble radius. Moreover, the following definitions were used:
\begin{eqnarray}
f(a) &=& \sqrt{\frac{\Omega_{m0}}{a} + \Omega_c(a)a^2},\\
\dot\phi(a) &=& \sqrt{3\Omega_{c0}}\sqrt{g(a)^{1/(1 + \alpha)} - \bar A g(a)^{-\alpha/(1 + \alpha)}},\\
V(a) &=& \frac{3}{2}\Omega_{c0}\biggr(g(a)^{1/(1 + \alpha)} + \bar A g(a)^{-\alpha/(1 + \alpha)}\biggl),\\
V_\phi(a) &=& \frac{f(a)}{\dot\phi}V'(a),\\
V_{\phi\phi}(a) &=& \frac{f(a)}{\dot\phi}V'_\phi(a),\\
\Omega_c(a) &=& \Omega_{c0}g(a)^{1/(1 + \alpha)}.
\end{eqnarray}
In these expressions, $\Omega_{c0}$ is the density parameter for the generalized Chaplygin gas model, which obeys the flat condition
$\Omega_{c0} + \Omega_{m0} = 1$ and $g(a)$ is given by (\ref{def}).
\par
\begin{center}
\begin{figure}[!t]
\begin{minipage}[t]{0.3\linewidth}
\includegraphics[width=\linewidth]{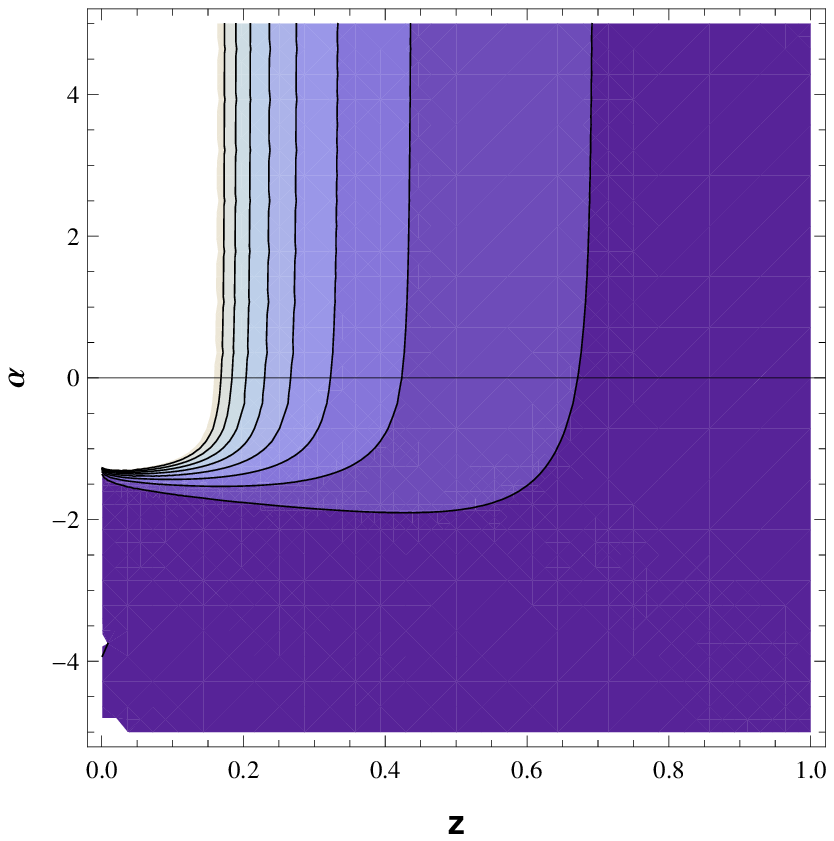}
\end{minipage} \hfill
\begin{minipage}[t]{0.3\linewidth}
\includegraphics[width=\linewidth]{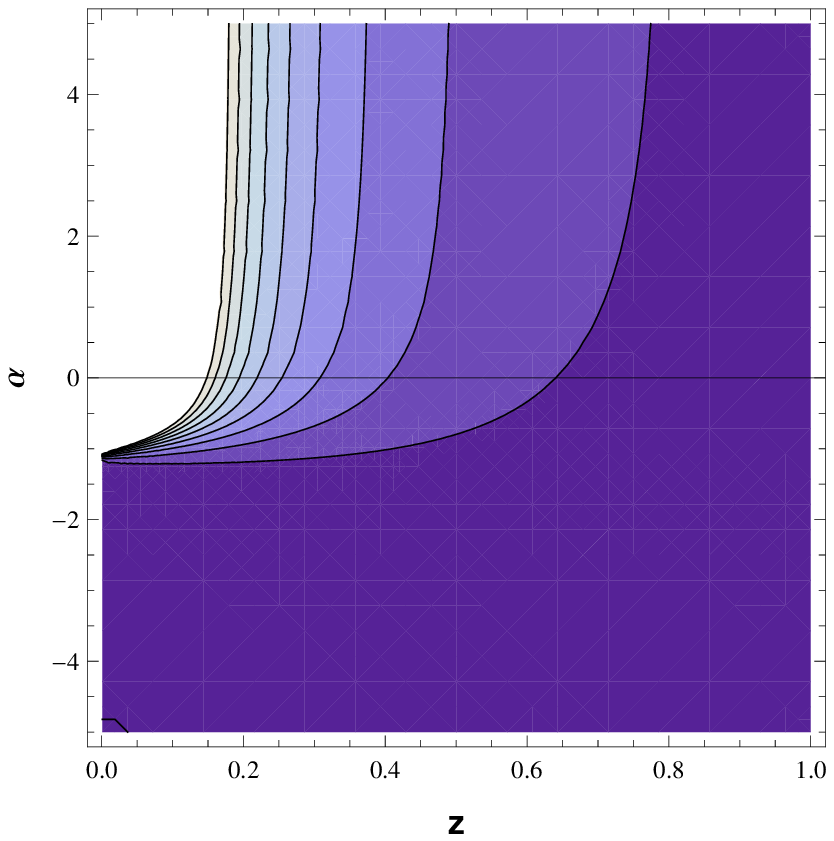}
\end{minipage} \hfill
\begin{minipage}[t]{0.3\linewidth}
\includegraphics[width=\linewidth]{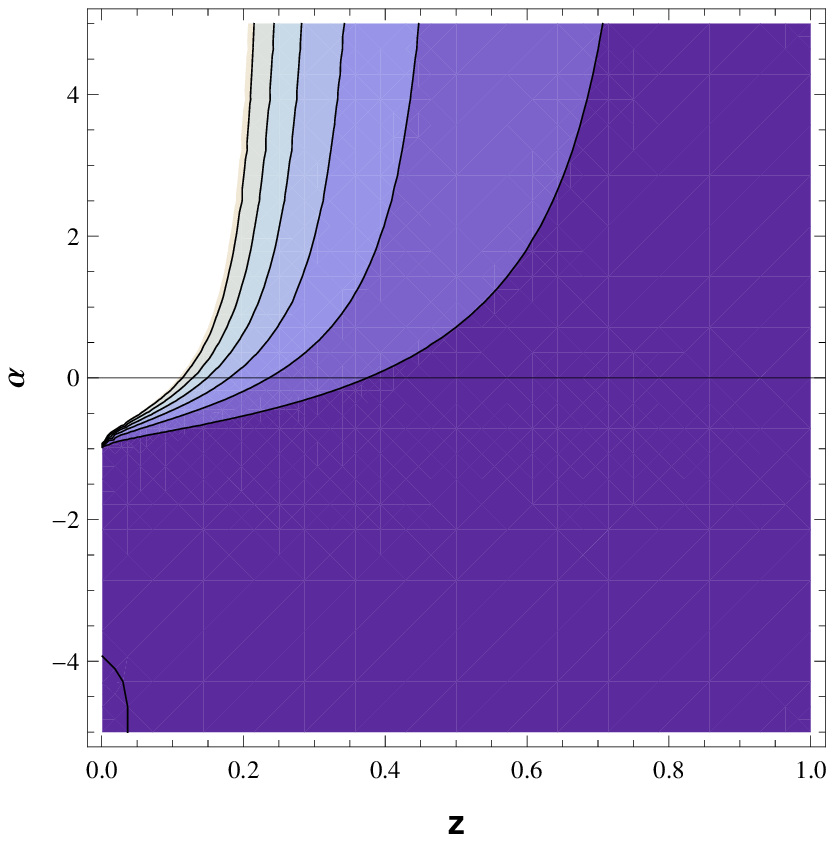}
\end{minipage} \hfill
\caption{{\protect\footnotesize Behavior of $\dot\phi$ as function of $\alpha$ and $z$, fixing $\Omega_{m0} = 0.3$ for
$\bar A =$ 0.1 (left), 0.5 (center) and 0.9 (right). The darker colors indicate an almost zero value for $\dot\phi$.
}}
\label{dotphi}
\end{figure}
\end{center}

Now, we compare the model with the power spectrum observational data from the 2dFGRS compilation in the range that corresponds to the linear regime, that
is $0.01\,Mpc^{-1} < kh^{-1} < 0.185\,Mpc^{-1}$. This compilation amounts to 39 data. In the numerical computation, one important aspect is how to
introduce the initial conditions. We use the BBKS transfer function \cite{bbks}, which gives the spectrum today for the $\Lambda$CDM model, integrating it
back to the redshift $z = 1000$, where the initial condition are fixed. The general procedure is described in reference \cite{sola}. In comparison with
this reference, there is an important modification. When $\alpha < - 1$, the general behavior is the dominance of an accelerated phase in the past,
approaching a dust dominated universe in recent times. This scenario may bring problems concerning the formation of structure (to which extent it
will be verified in what follows), but may be interesting to implement the idea that the acceleration of the universe is a transient phase. In fact,
in reference \cite{staro}, it has been pointed out that the acceleration of the universe is slowing down. The important point is that, when 
$\alpha \stackrel{<}{\sim} - 3$, the resulting scenario is very similar to the $\Lambda$CDM model until very recent times: for $\alpha \sim - 3$ 
it is found $\dot\phi \sim 0$ and $V(\phi) =$ constant until $z \sim 20$. The general behavior of $\dot\phi$ in terms of $z$ and $\alpha$ is
illustrated, for some specific cases, in figure (\ref{dotphi}). For $\alpha > - 1$ this general behavior is inverted. The fact that for $\alpha$ negative enough
the model is essentially $\Lambda$CDM until very recently does not allow to imposed the initial conditions using the transfer function
with $\Omega_{m0} \sim 0.3$ and $\Omega_{\Lambda0} = 0.7$ as it has been done in \cite{sola}.
For $\alpha  \stackrel{<}{\sim} - 3$ the initial conditions must be imposed,
due to computational reasons, about $z = 20$. Because of that, we must take into account the fact that
the transfer function depends explicitly on the mass parameters. 

\par
The fundamental quantity to be evaluated is the matter power spectrum
\begin{equation}
P_k = \delta_k^2,
\end{equation}
which is the Fourier transform of the two-points correlation function of matter distribution in the universe.
We perform a Bayesian statistical analysis, using first the $\chi^2$ function
\begin{equation}
\chi^2 = \sum_{i=1}^n \frac{(P_{k_i}^o - P_{k_i}^t)^2}{\sigma_i^2},
\end{equation}
where $P_{k_i}^o$ is the $i^{th}$ power spectrum observational data, with $\sigma_i^2$ observational error bar, $P_{k_i}^t$ its corresponding theoretical prediction. From this quantity, we construct the Probability Distribution Function (PDF),
\begin{equation}
P = A e^{- \chi^2/2},
\end{equation}
where $A$ is a normalization constant. The PDF depends in general, for the GCG model, of 5 parameters: $\bar A$, $\alpha$, $\Omega_{dm0}$
$\Omega_{c0}$ and $\Omega_{k0}$, where the last quantity is associated with the curvature of the spatial section. The power spectrum is
expressed as function of $h$, reducing the number of parameters to four. Imposing the flat condition $\Omega_{k0} = 0$, implies only
3 free parameters. We will perform a two-dimensional computation, fixing $\bar A$. First, the computations will be made considering three
values of this quantity: $\bar A = 0.1, 0.5$ and $0.9$. This will allow us to evaluate the influence of this parameter in the final results.
We will plot the two dimensional PDF as well as the corresponding one dimensional PDF, by marginalizing (integrating) over the one of the variables.
\par
\begin{center}
\begin{figure}[!t]
\begin{minipage}[t]{0.3\linewidth}
\includegraphics[width=\linewidth]{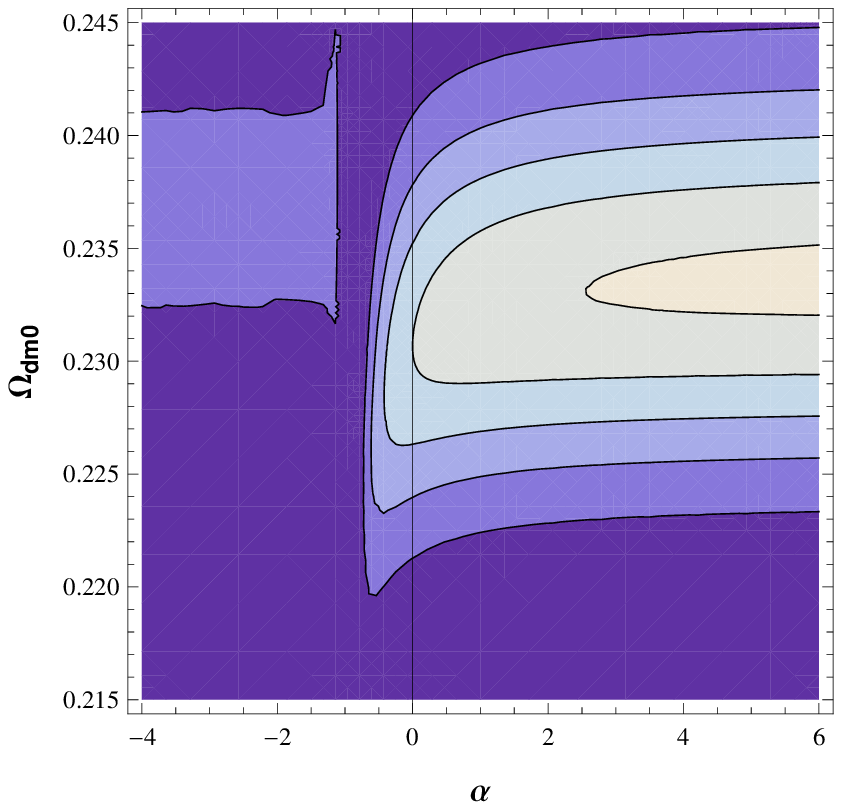}
\end{minipage} \hfill
\begin{minipage}[t]{0.3\linewidth}
\includegraphics[width=\linewidth]{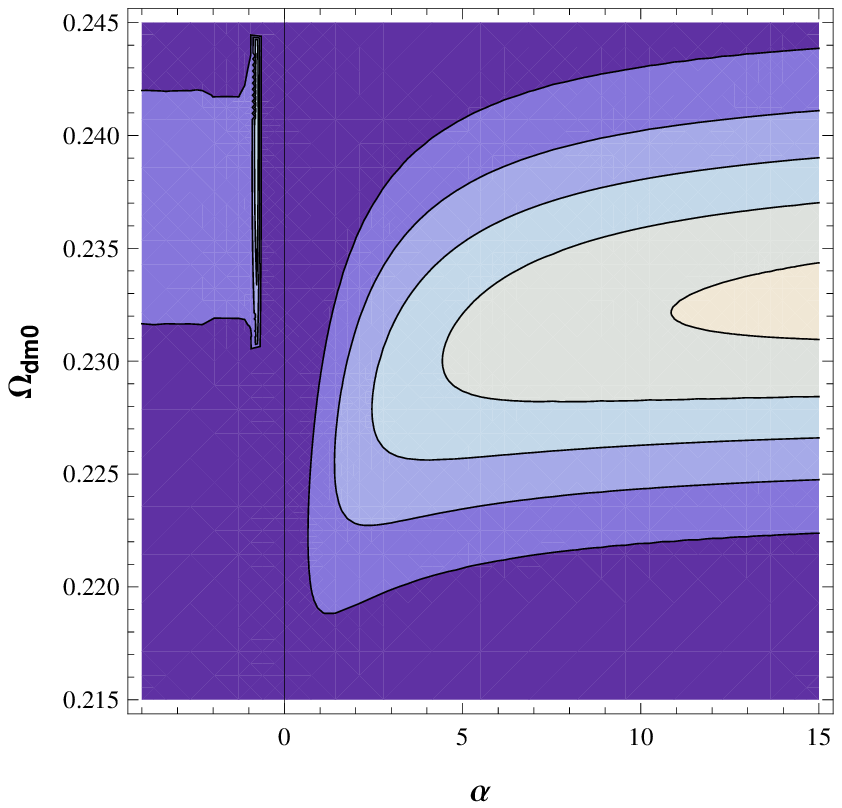}
\end{minipage} \hfill
\begin{minipage}[t]{0.3\linewidth}
\includegraphics[width=\linewidth]{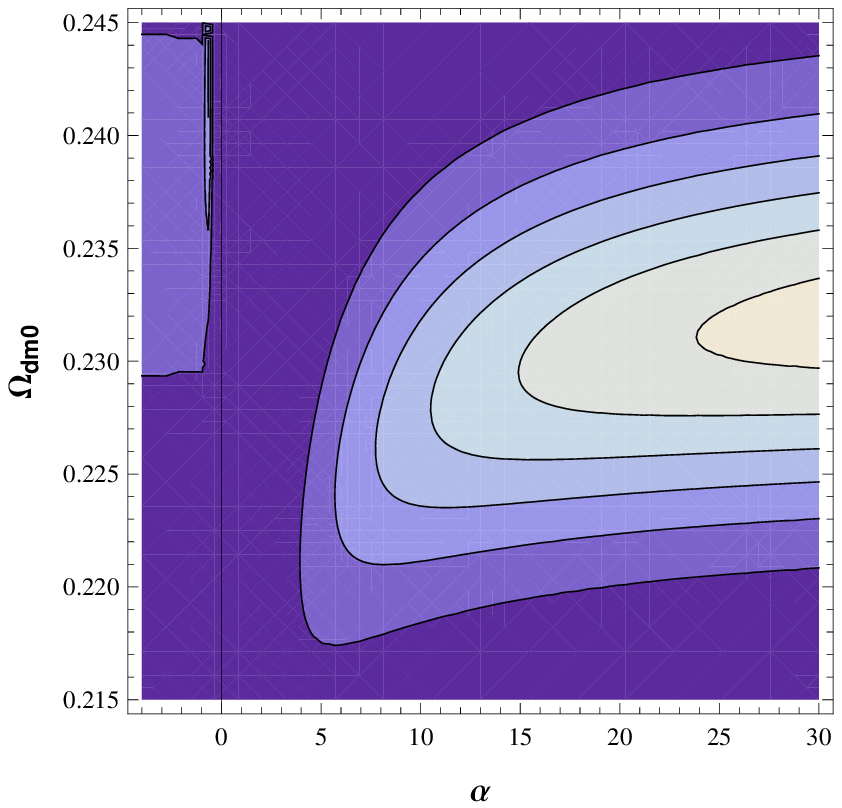}
\end{minipage} \hfill
\caption{{\protect\footnotesize Two dimensional PDF for the parameters $\Omega_{dm0}$ and $\alpha$ for $\bar A = 0.1$ (left), 0.5 (center) and 0.9 right.}}
\label{omega-alpha}
\end{figure}
\end{center}
\begin{center}
\begin{figure}[!t]
\begin{minipage}[t]{0.3\linewidth}
\includegraphics[width=\linewidth]{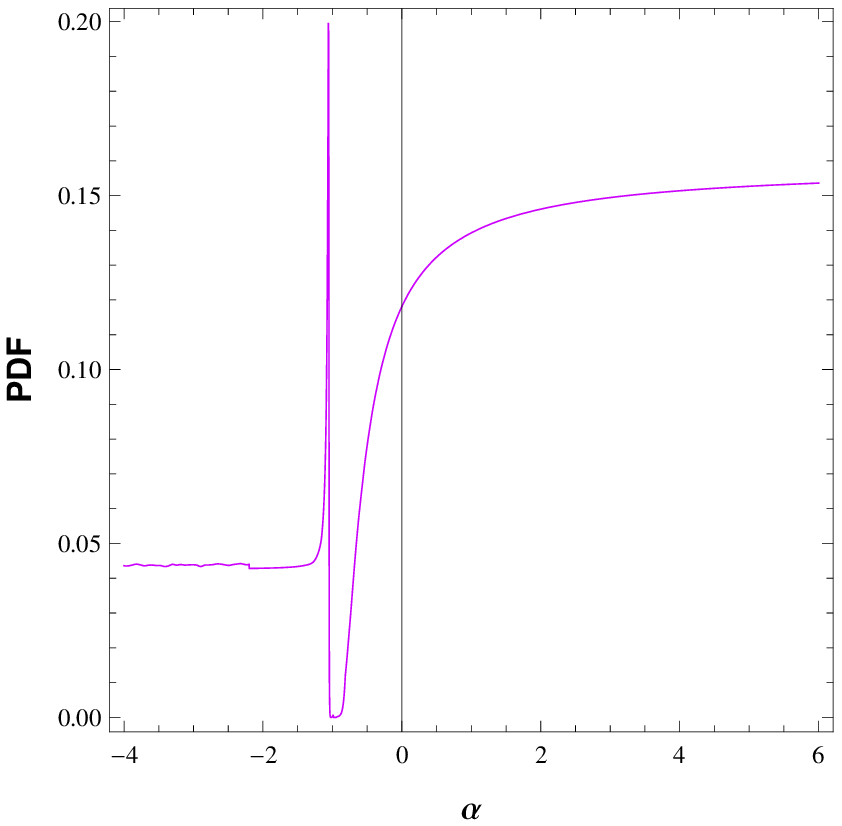}
\end{minipage} \hfill
\begin{minipage}[t]{0.3\linewidth}
\includegraphics[width=\linewidth]{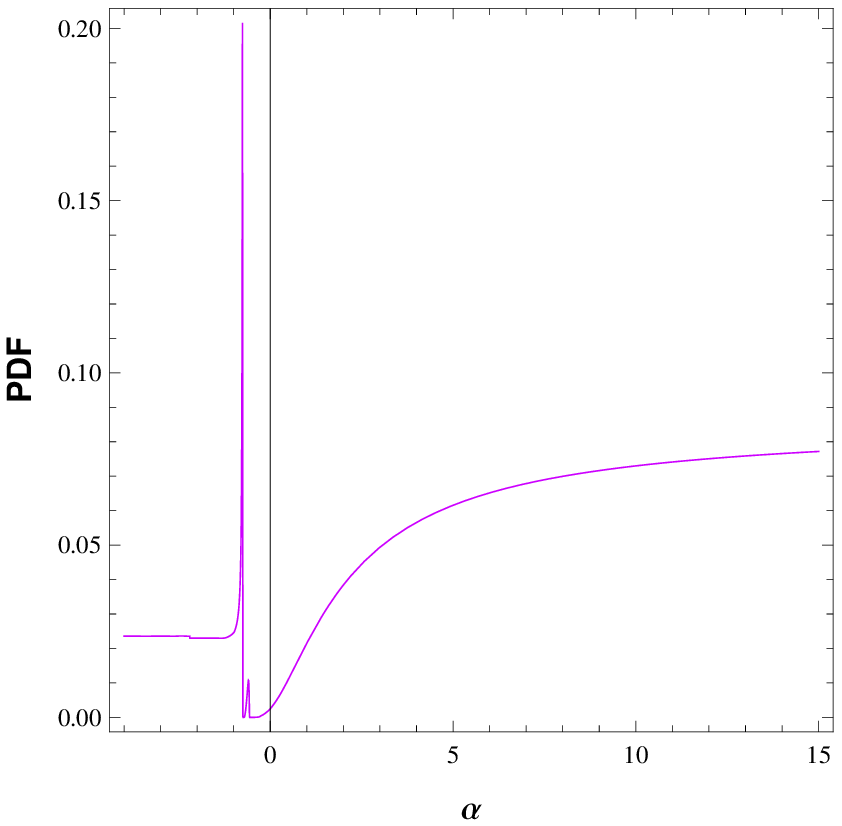}
\end{minipage} \hfill
\begin{minipage}[t]{0.3\linewidth}
\includegraphics[width=\linewidth]{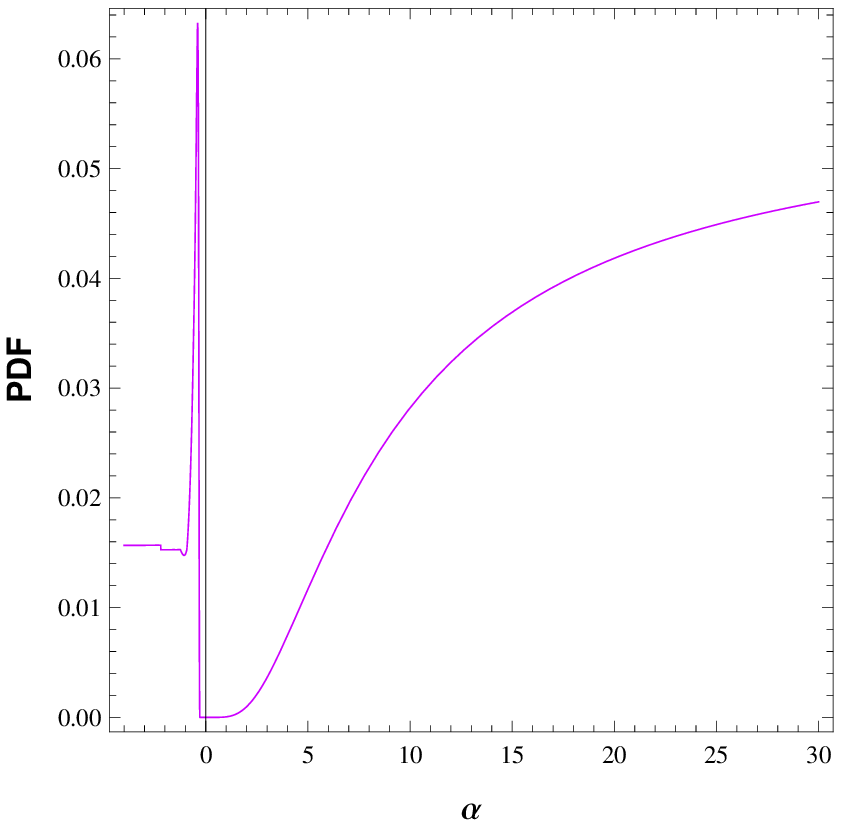}
\end{minipage} \hfill
\caption{{\protect\footnotesize One dimensional PDF for the parameter $\alpha$ for $\bar A = 0.1$ (left), 0.5 (center) and 0.9 right.
}}
\label{alpha}
\end{figure}
\end{center}
\begin{center}
\begin{figure}[!t]
\begin{minipage}[t]{0.3\linewidth}
\includegraphics[width=\linewidth]{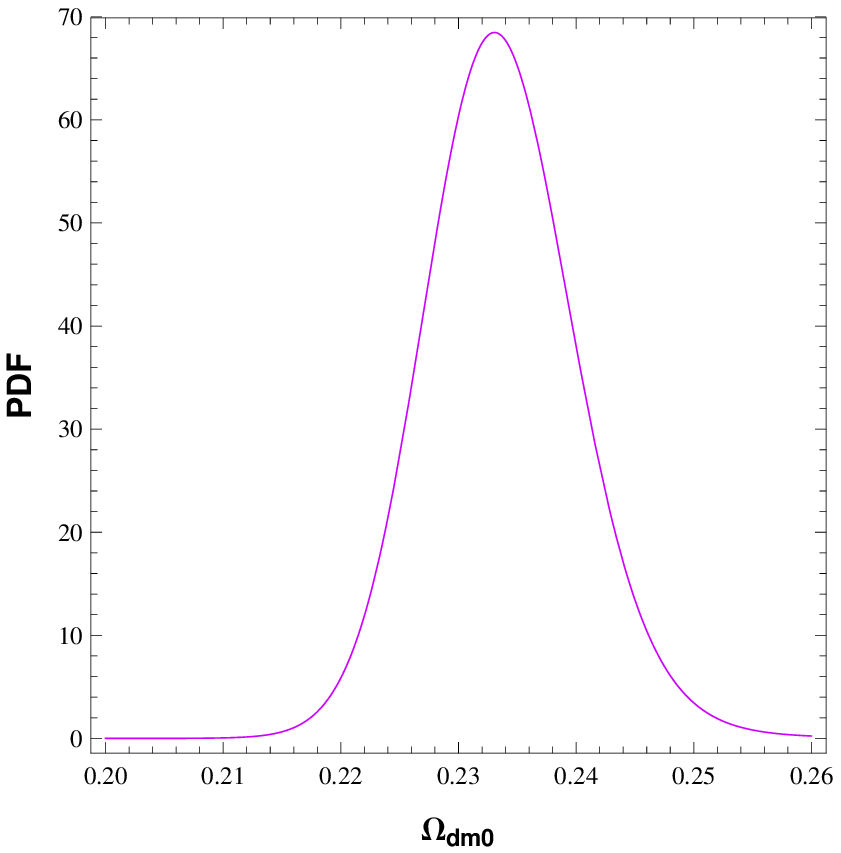}
\end{minipage} \hfill
\begin{minipage}[t]{0.3\linewidth}
\includegraphics[width=\linewidth]{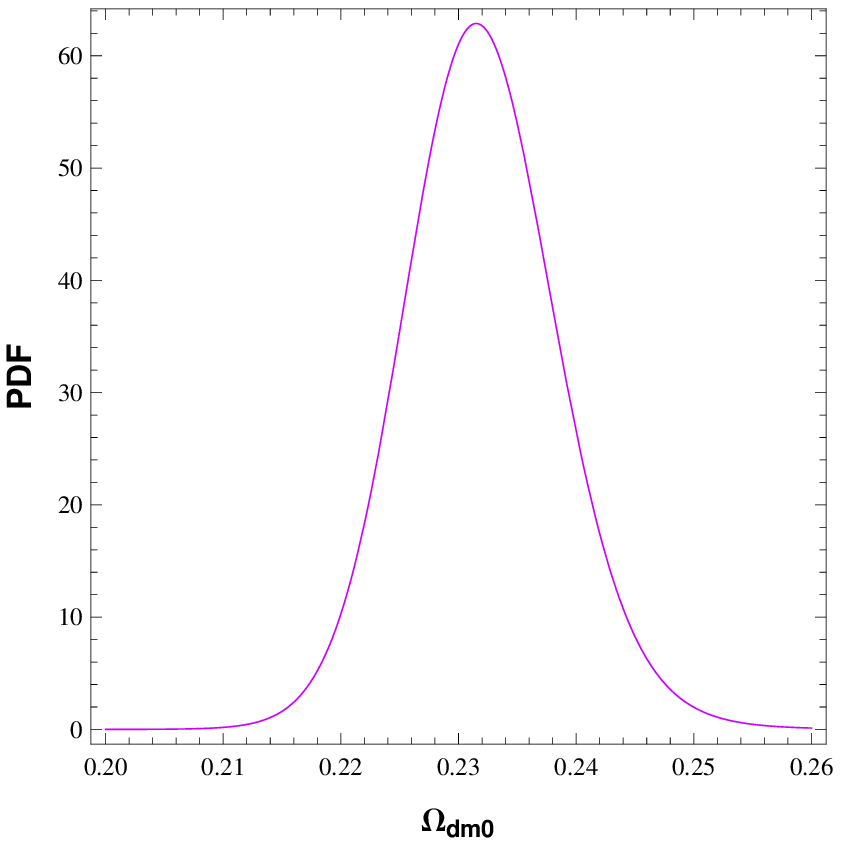}
\end{minipage} \hfill
\begin{minipage}[t]{0.3\linewidth}
\includegraphics[width=\linewidth]{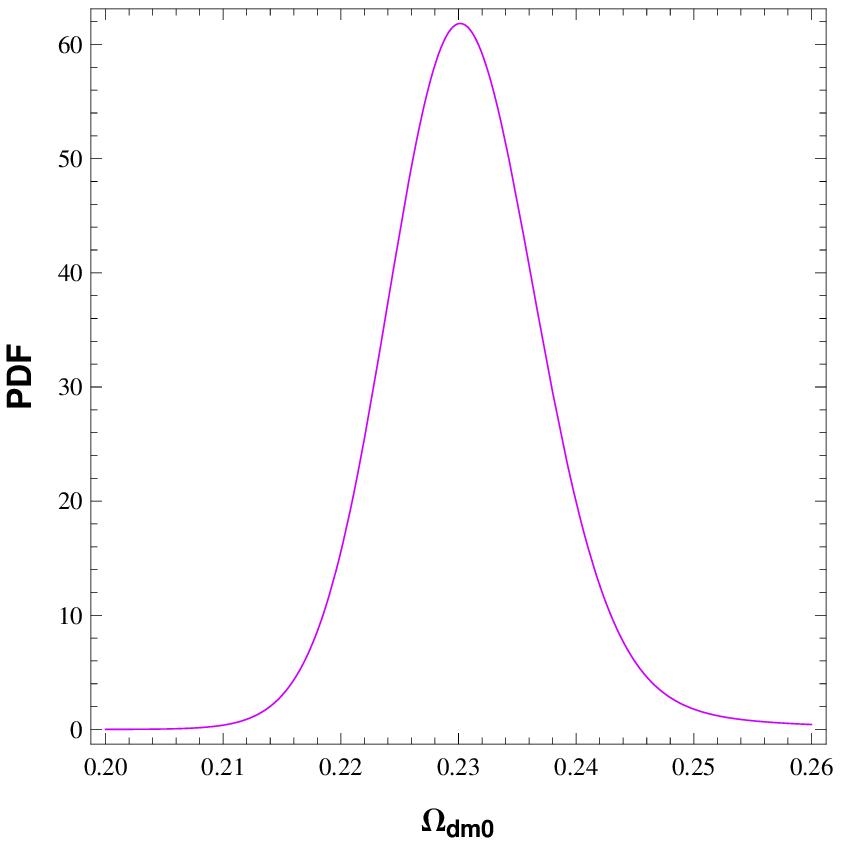}
\end{minipage} \hfill
\caption{{\protect\footnotesize One dimensional PDF for the parameter $\Omega_{dm0}$ for $\bar A = 0.1$ (left), 0.5 (center) and 0.9 right.
}}
\label{omega}
\end{figure}
\end{center}

The results indicate a very general pattern. Leaving $\alpha$ free, we remark the appearance of the two plateau in the PDF: one for $\alpha \stackrel{<}{\sim} - 2.5$, and the other for $\alpha \stackrel{>}{\sim} 1$. The plateau corresponding to the positive values of $\alpha$ is higher
than the plateau corresponding to negative values. Hence, the model predicts, in some sense, that $\alpha$ must be positive. From figures (\ref{omega-alpha},\ref{alpha},\ref{omega}) it is clear that the PDF is concentrated in the region of positive values for $\alpha$ and values for $\Omega_{dm0}$
in the range $0.2 < \Omega_{dm0} < 0.3$. These regions are displaced for larger positive values of $\alpha$ as $\bar A$ increases. There is
another region with smaller probabilities for $\alpha$ negative - see figures (\ref{omega-alpha},\ref{alpha},\ref{omega}) - which is almost not affected by changing
$\bar A$. These behaviors are confirmed by considering the one dimensional PDF for $\alpha$ displayed in figure
(\ref{alpha}). Remark that there is a peak near $\alpha = - 1$ which corresponds to a linear, constant, relation between pressure and density. 
We may attribute this feature to effects corresponding to the numerical computation near the singular point - however, even increasing strongly the
precision of the computation (implying increasing the computational time), this peak does
not disappear.
\par
For the dark matter density parameter $\Omega_{dm0}$, the PDF is peaked quite generally around $\Omega_{dm0} = 0.23$. This is a result very similar to
the $\Lambda$CDM model, and all framework is very close to the quintessence model (whose precise prediction depends on the choice of the potential
for the scalar field). The parameter estimations at, for example, $1\sigma$, $2\sigma$, etc. for the parameter $\alpha$ becomes very doubtful due to the existence of the
two plateaus. It could be done for the parameter $\Omega_{dm0}$ due to its almost gaussian probability distribution, but even a visual analysis
of figures (\ref{omega}) shows that they change very little with the increasing value of $\bar A$, remaining always peaked around $0.23$.

\par
\begin{center}
\begin{figure}[!t]
\begin{minipage}[t]{0.3\linewidth}
\includegraphics[width=\linewidth]{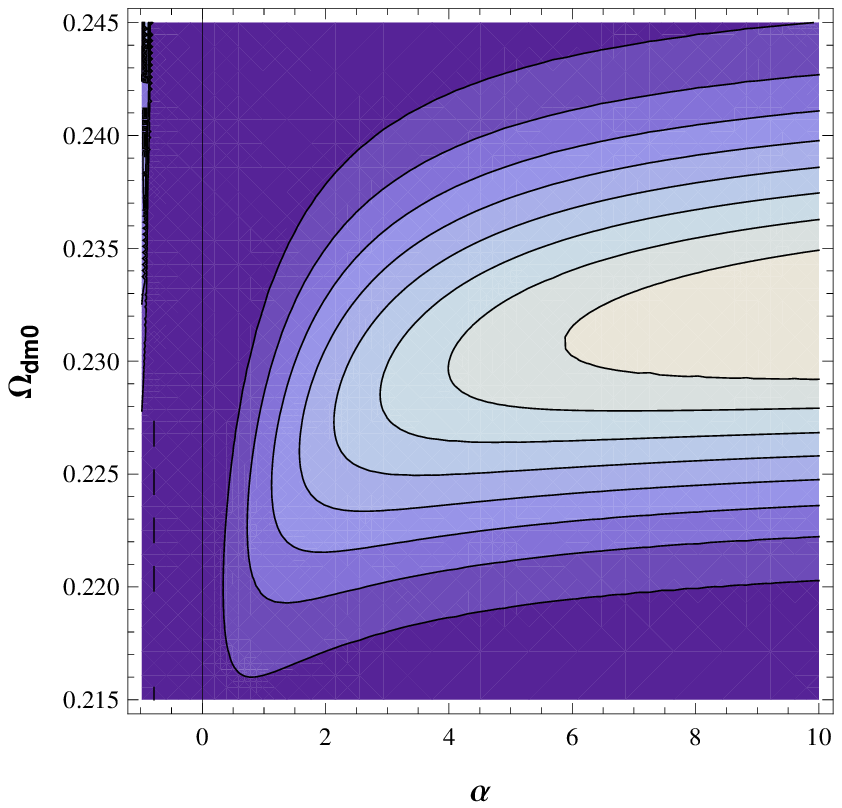}
\end{minipage} \hfill
\begin{minipage}[t]{0.3\linewidth}
\includegraphics[width=\linewidth]{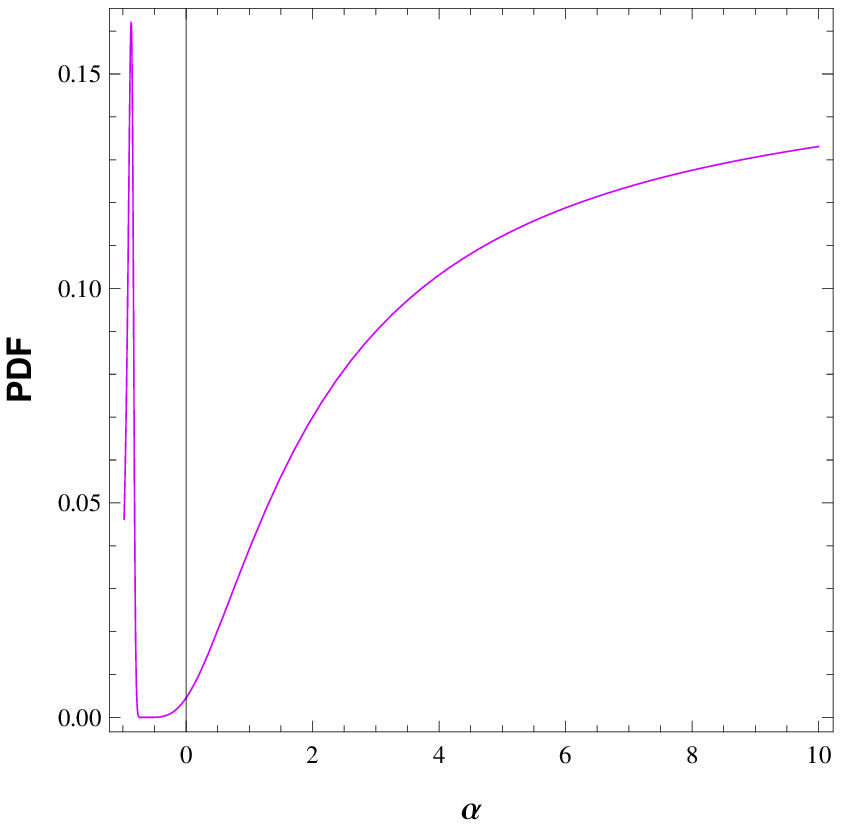}
\end{minipage} \hfill
\begin{minipage}[t]{0.3\linewidth}
\includegraphics[width=\linewidth]{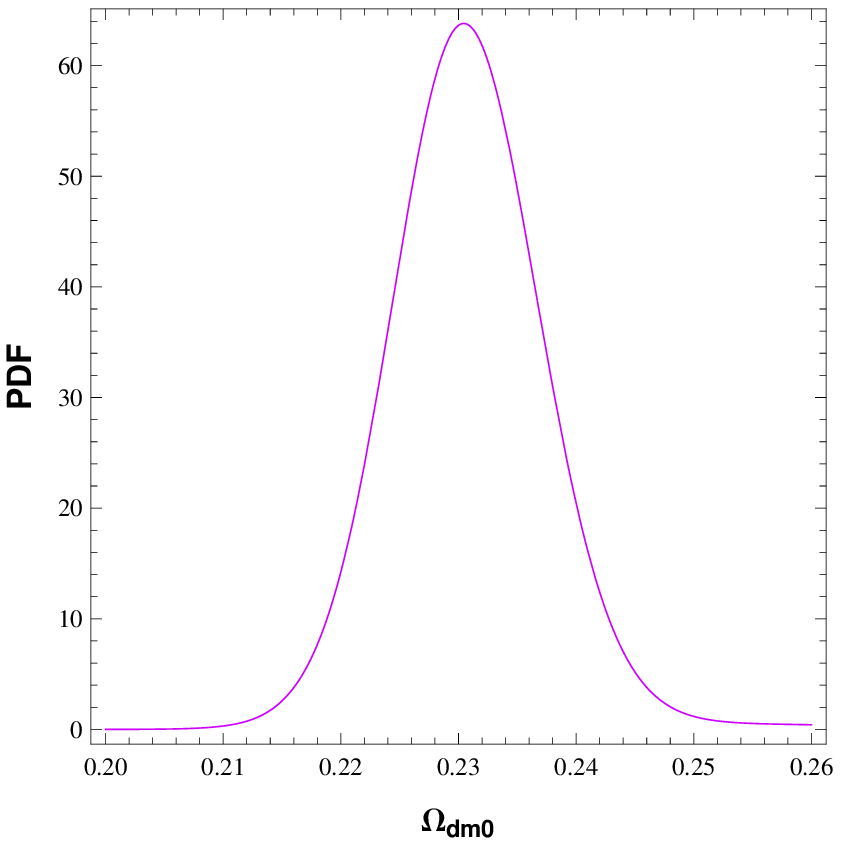}
\end{minipage} \hfill
\caption{{\protect\footnotesize Two and one dimensional PDFs restricting $\alpha > - 1$ and fixing $\bar A = 0.5$.
}}
\label{-1}
\end{figure}
\end{center}
\begin{center}
\begin{figure}[!t]
\begin{minipage}[t]{0.3\linewidth}
\includegraphics[width=\linewidth]{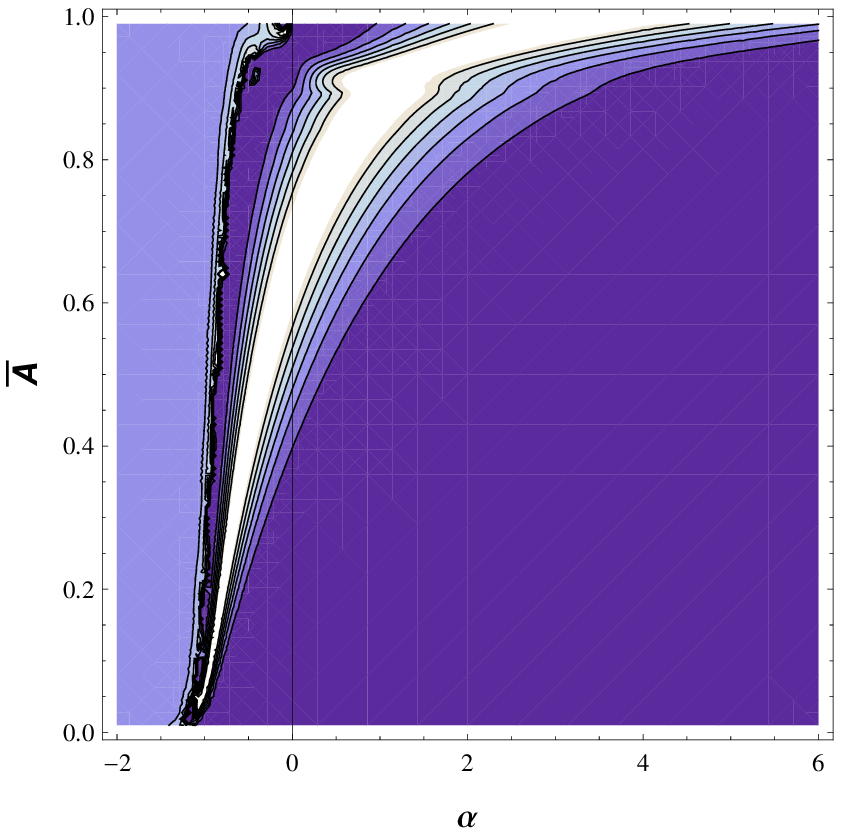}
\end{minipage} \hfill
\begin{minipage}[t]{0.3\linewidth}
\includegraphics[width=\linewidth]{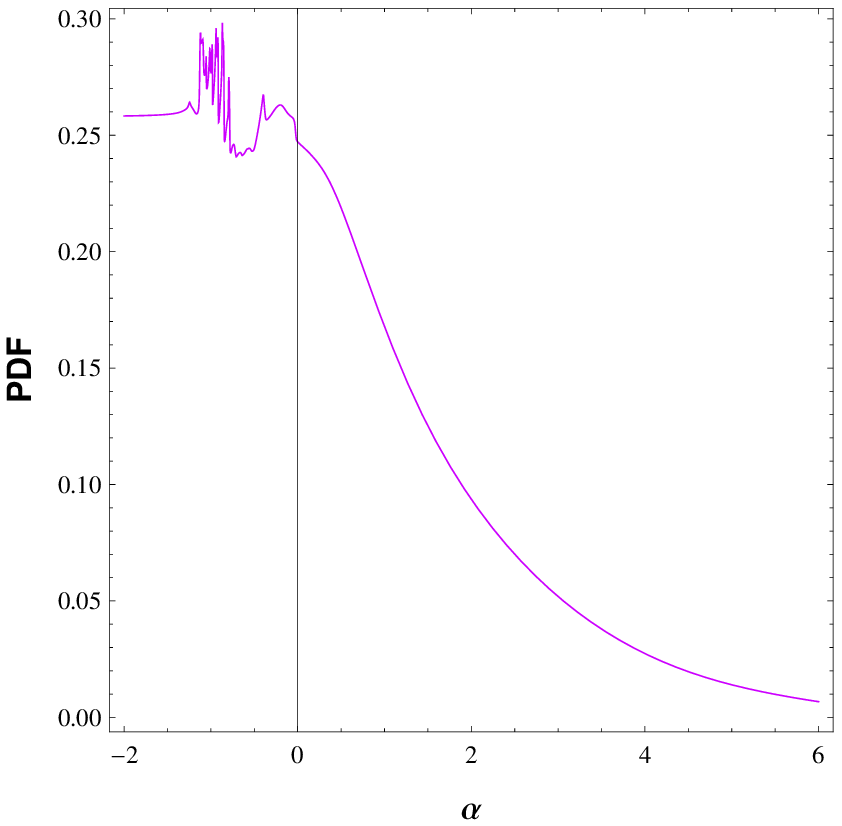}
\end{minipage} \hfill
\begin{minipage}[t]{0.3\linewidth}
\includegraphics[width=\linewidth]{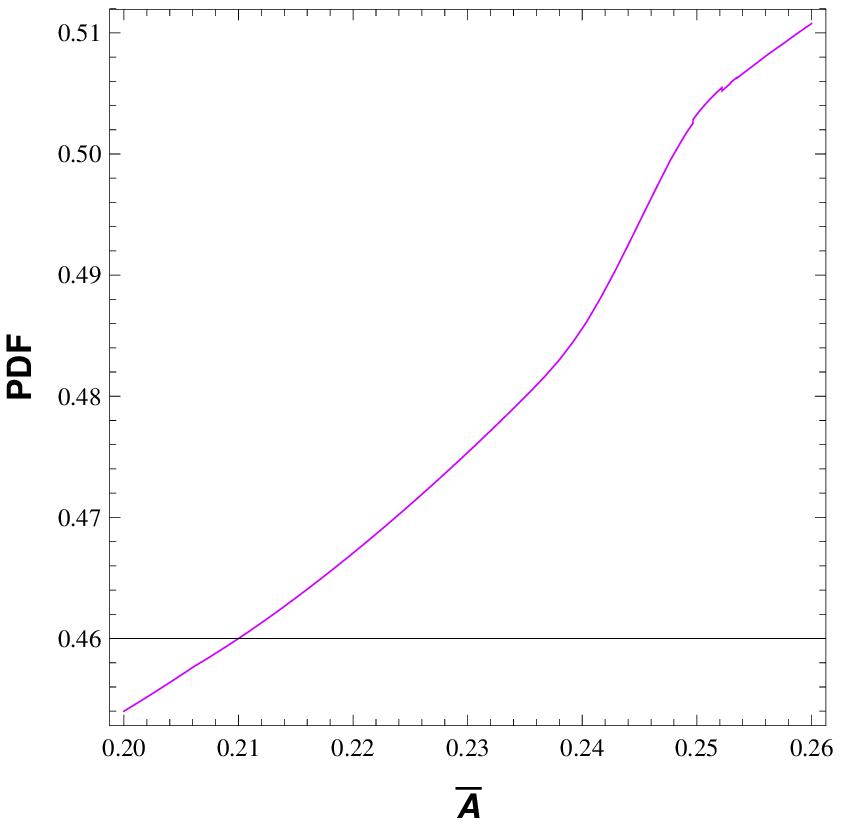}
\end{minipage} \hfill
\caption{{\protect\footnotesize Two and one dimensional PDFs restricting $\Omega_{dm0} = 0$.
}}
\label{unif}
\end{figure}
\end{center}

It can be expected that the results may change by imposing some prior. One possibility is to restrict $\alpha > - 1$. This restriction does
not change the results, which are displayed in figure (\ref{-1}). Again, the maximum of probability for $\Omega_{dm0}$ is sharply peaked around
$0.23$, while there is a plateau for the PDF of $\alpha$ which, in this case, by fixing $\bar A = 0.5$, begins at $\alpha \sim 8$. However, some
interesting effects appear if we fix, from the beginning, the dark matter component equal to zero. This amounts to impose the unification scenario
where the dark matter and dark energy are both represented by the Chaplygin gas model. In this case, two free parameters are
considered: $\bar A$ and $\alpha$. In figure (\ref{unif}) the results are displayed. The main feature to be remarked is that, now the
maximum for PDF occurs for negative $\alpha$. There are strong oscillations near $\alpha = - 1$. In this case,
the minimum for $\chi^2$ may depend on the precision of the numerical evaluation. It can vary from, for example, $\chi^2 \sim 0.30$ for
$\alpha \sim -0.85$, to $\chi^2 = 0.34$ for $\alpha = - 0.88$ (using SN Ia, Gold sample, the favored value is
$\alpha = -0.10$\cite{colistete}). When the unification scenario is not imposed from the beginning (the previous cases), it is found
typically $\chi^2_{min} = 0.30$, in a more stable way. For a comparison, for the $\Lambda$CDM, we have
$\chi^2_{min} \sim 0.38$, but with just one free parameter when the spatial section is flat. If the Akaike Information Criteria, $AIC$, is
used, which allows to compare models with different number of free parameters ($AIC = 2\times N + \chi^2_{total}$, where $N$ is
the number of free parameters), the $\Lambda$CDM model remains the best one, but the difference
is very small: $AIC \sim 18$ for the scalar GCG model with three free parameters, $AIC \sim 17-18$, for the unified scalar GCG model with two free parameters,
while $AIC \sim 17$ for the $\Lambda$CDM model.

\section{The Rastall's unification scalar model}

The results described above are interesting but suffer from a major drawback: a canonical self-interacting scalar model hardly can represent dark matter, since
its sound speed is different from zero, in fact it is equal to 1. One of the main properties of dark matter is a zero sound speed already in the radiation era (at least in the end of the radiation era). Hence, the previous model has to be seen as an effective model valid during the matter dominated era, which must be
complemented for previous era. However, there is another possibility to describe a dark
matter scalar model, or even a dark matter/dark energy scalar model, which is a particular case of the scalar-tensor description of the Rastall's theory of gravity. This possibility is explored in some detail in reference \cite{liddle}. Let us follows this approach now.
\par
Rastall's theory has been proposed in the beginning of the 70's in order to take into account a fundamental idea: the usual conservation law for
the energy-momentum tensor has been tested only in flat space-time, and in curved space-time its employment is a pure extrapolation. Hence,
Rastall has proposed that the Einstein's equations should be written as
\begin{equation}
R_{\mu\nu} - \frac{\lambda}{2}g_{\mu\nu}R = 8\pi GT_{\mu\nu},
\end{equation}
where $\lambda$ is a free parameter. If $\lambda = 1$, General Relativity is recovered. The Rastall's equations may be re-written
as,
\begin{eqnarray}
R_{\mu\nu} - \frac{1}{2}g_{\mu\nu}R &=& 8\pi G\biggr[T_{\mu\nu} - \frac{2 - \gamma}{2}g_{\mu\nu}T\biggl],\\
{T^{\mu\nu}}_{;\mu} &=& \frac{\gamma -1}{2}T^{;\nu}, \quad \gamma = \frac{3\lambda - 2}{2\lambda - 1}
\end{eqnarray}
One of the main theoretical drawbacks of the Rastall's theory is the absence of a Lagrangian formulation. However, a possible action principle has been proposed
in reference \cite{smalley}, but outside the framework of riemannian geometry. 
In reference \cite{hamani} Rastall's theory has been tested against the 2dFGRS data for power spectrum. In order to do so, it has been considered
a two-fluid model, one obeying the usual conservation law, and the other the Rastall's conservation law. This hybrid model was necessary in order
to represent the baryon component, which must have zero pressure, clustering to form the observed structures. The second component should represent dark energy, and can have a zero pressure but behaving effectively as a component with negative pressure. The hydrodynamical representation leads to
a very stringent limit on the parameter $\gamma$, reducing it effectively to the Einstein's theory.
\par
Hence, a scalar model was conceived in order
to represent the "Rastall's fluid". In this case, the equations read now,
\begin{eqnarray}
\label{em}
 R_{\mu\nu} - \frac{1}{2}g_{\mu\nu}R = 8\pi GT_{\mu\nu} + \phi_{;\mu}\phi_{;\nu} - \frac{2 - \gamma}{2}g_{\mu\nu}\phi_{;\rho}\phi^{;\rho} + g_{\mu\nu}(3 - 2\gamma )V(\phi),\\
 \label{kgm}
 \nabla_\rho\nabla^\rho\phi + (3 - 2\gamma)V_\phi = (1 - \gamma)\frac{\phi^{;\rho}\phi^{;\sigma}\phi_{;\rho;\sigma}}{\phi_{;\alpha}\phi^{;\alpha}}.
\end{eqnarray}
As before, $T_{\mu\nu}$ represents the baryons and obeys the usual conservation law. Again, the problem of a action principle leading to
equations (\ref{em},\ref{kgm}) must be considered, and it can be treated in a similar way as the formulation of reference \cite{smalley}. It is worth to mention that
the modified Klein-Gordon equation (\ref{kgm}) is similar to some terms appearing in the Galileon's theory \cite{galileon}, but it seems that no exact correspondence
can be established between the scalar Rastall model and the Galileon framework.
\par
The interesting point about this model is that the Rastall's self-interacting scalar field can have a zero pressure behavior at perturbative level
when $\gamma = 2$. This has been point out in reference \cite{liddle}, without an explicit mention to the Rastall's theory.
\par
Hence, to have a zero sound velocity, the scalar model must obey the following equations:
\begin{eqnarray}
 R_{\mu\nu} - \frac{1}{2}g_{\mu\nu}R = 8\pi GT_{\mu\nu} + \phi_{;\mu}\phi_{;\nu} + g_{\mu\nu}V(\phi),\\
 \nabla_\rho\nabla^\rho\phi + V_\phi + \frac{\phi^{;\rho}\phi^{;\sigma}\phi_{;\rho;\sigma}}{\phi_{;\alpha}\phi^{;\alpha}} &=& 0,
\end{eqnarray}
where, just for future convenience, we have made the redefinition $V(\phi) \rightarrow - V(\phi)$. In this case, the energy-momentum tensor
for the scalar field is given by,
\begin{equation}
T_{\mu\nu}^\phi = \phi_{;\mu}\phi_{;\nu} + g_{\mu\nu}V(\phi).
\end{equation}
Hence, the energy and the density of the scalar field is given by the following expressions:
\begin{eqnarray}
\rho_\phi &=& \dot\phi^2 + V(\phi),\\
p_\phi &=& - V(\phi).
\end{eqnarray}
Now, we must to proceed as in the previous section to fix $\dot\phi$ and $V(\phi)$.
Imposing that the scalar field must represent the GCG, we find
\begin{eqnarray}
\dot\phi(a) &=& \sqrt{3\Omega_{c0}}\sqrt{g(a)^{1/(1 + \alpha)} - \bar A g(a)^{-\alpha/(1 + \alpha)}},\\
V(a) &=& 3\Omega_{c0}\bar A g(a)^{-\alpha/(1 + \alpha)}.
\end{eqnarray}
\par
The perturbed equations read now,
\begin{eqnarray}
\label{rastall1}
 \ddot\delta + 2\frac{\dot a}{a}\dot\delta - \frac{3}{2}\frac{\Omega_0}{a^3}\delta &=&
\dot\phi\dot\Psi - V_\phi\Psi,\\
\label{rastall2}
2\ddot\Psi + 3\frac{\dot a}{a}\dot\Psi + \biggr\{\frac{k^2}{a^2} + V_{\phi\phi}\biggl\}\Psi &=& \dot\phi\dot\delta,
\end{eqnarray}
where $\Psi = \delta\phi$ and $\delta$ is the density contrast of the matter component, as before.
Using now the scale factor as the variable, the above system of equations take the
following form:
\begin{eqnarray}
 \delta'' + \biggr\{\frac{2}{a} + \frac{f'(a)}{f(a)}\biggl\} \dot\delta - \frac{3}{2}\frac{\Omega_0}{a^3f^2(a)}\delta &=&
\phi'\Psi' - \frac{V_\phi}{f^2(a)}\Psi,\\
2\Psi'' + \biggr\{\frac{3}{a} + 2\frac{f'(a)}{f(a)}\biggl\}\Psi' + \biggr\{\frac{k^2}{a^2f^2(a)} + \frac{V_{\phi\phi}}{f^2(a)}\biggl\}\Psi &=& \phi'\delta',
\end{eqnarray}
The function $f$ has the same form as in the preceding section, but now $\dot\phi$ and $V(\phi)$ are given by (\ref{rastall1},\ref{rastall2}).
\par
\begin{center}
\begin{figure}[!t]
\begin{minipage}[t]{0.3\linewidth}
\includegraphics[width=\linewidth]{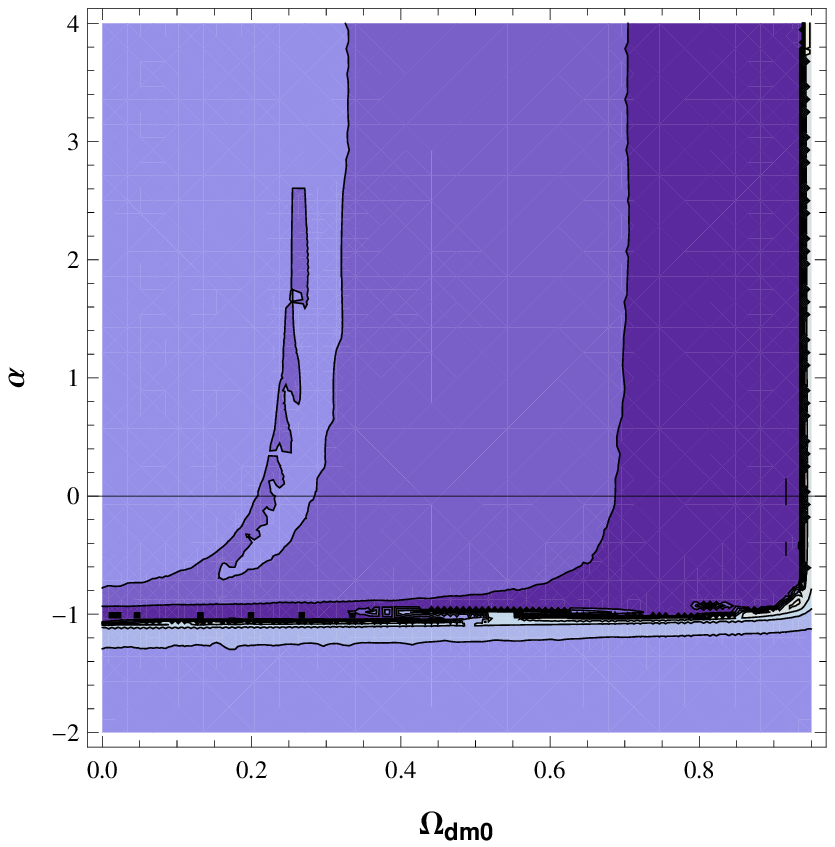}
\end{minipage} \hfill
\begin{minipage}[t]{0.3\linewidth}
\includegraphics[width=\linewidth]{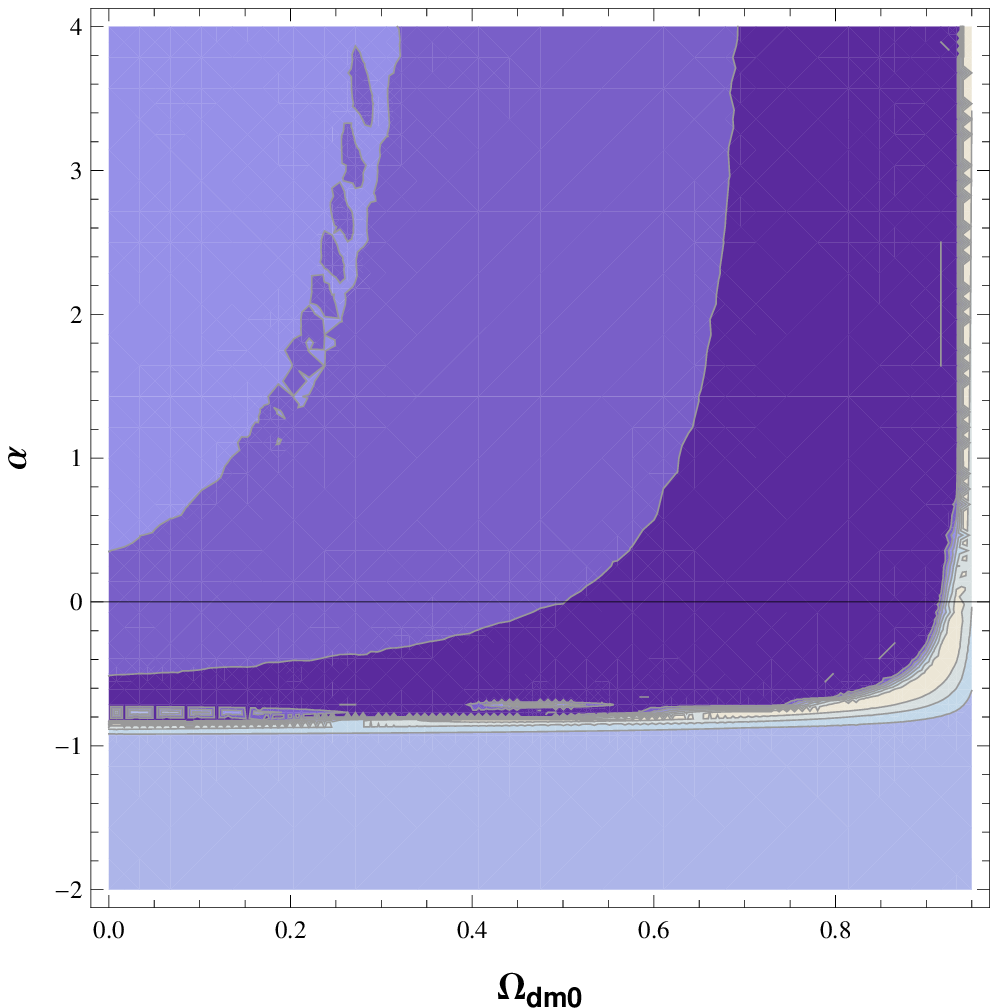}
\end{minipage} \hfill
\begin{minipage}[t]{0.3\linewidth}
\includegraphics[width=\linewidth]{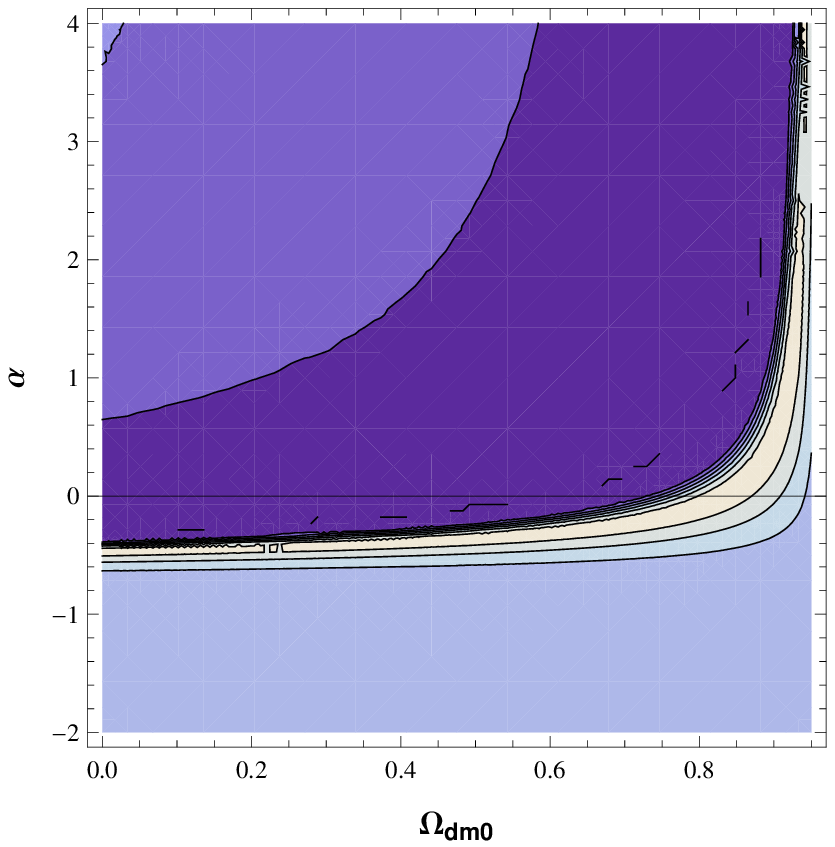}
\end{minipage} \hfill
\caption{{\protect\footnotesize Two dimensional PDFs for the Rastall's scalar model with $\bar A = 0.1$, $\bar A = 0.5$ and $\bar A = 0.9$. 
}}
\label{rastall-fig1}
\end{figure}
\end{center}
\begin{center}
\begin{figure}[!t]
\begin{minipage}[t]{0.3\linewidth}
\includegraphics[width=\linewidth]{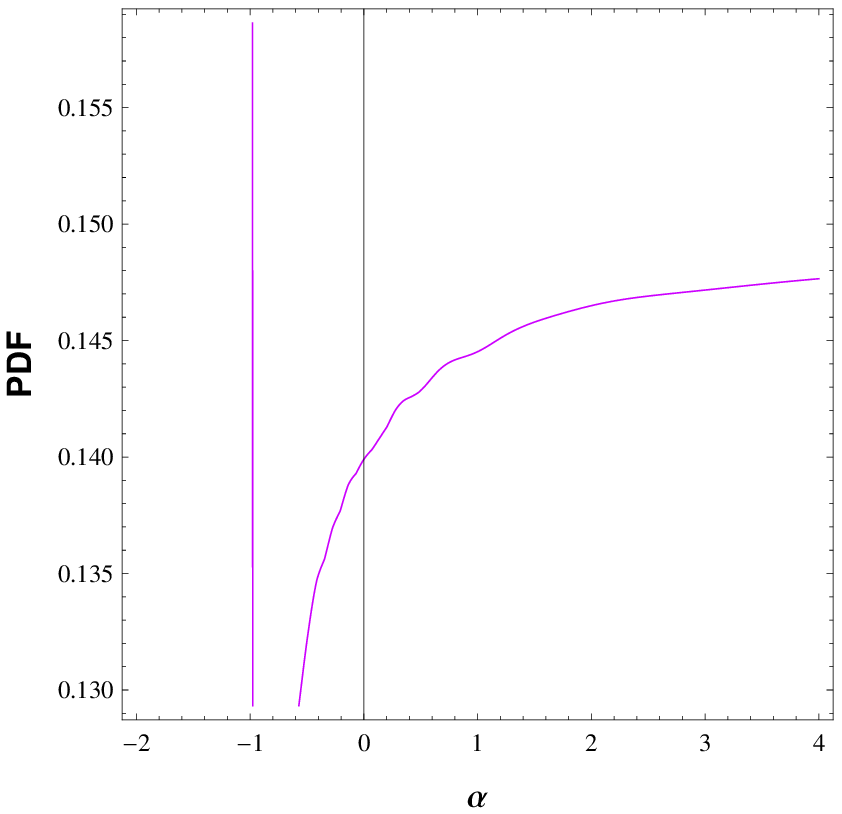}
\end{minipage} \hfill
\begin{minipage}[t]{0.3\linewidth}
\includegraphics[width=\linewidth]{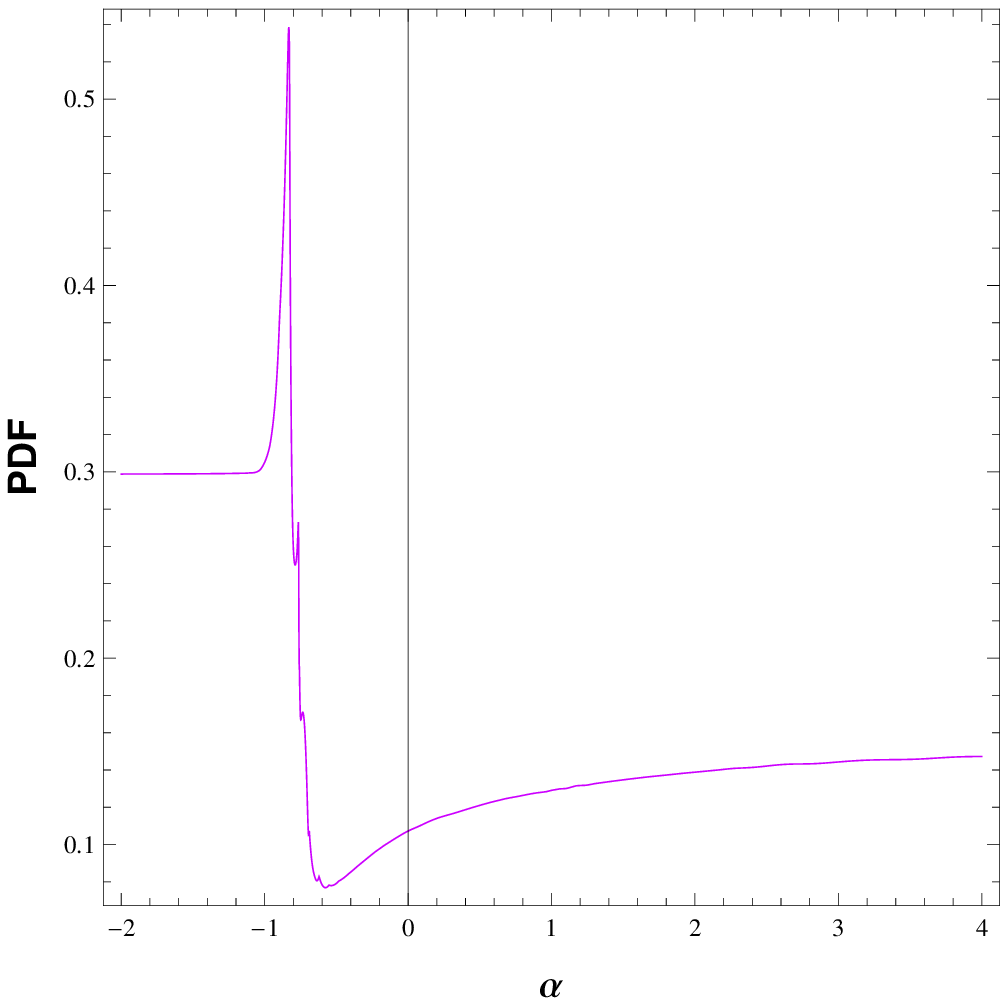}
\end{minipage} \hfill
\begin{minipage}[t]{0.3\linewidth}
\includegraphics[width=\linewidth]{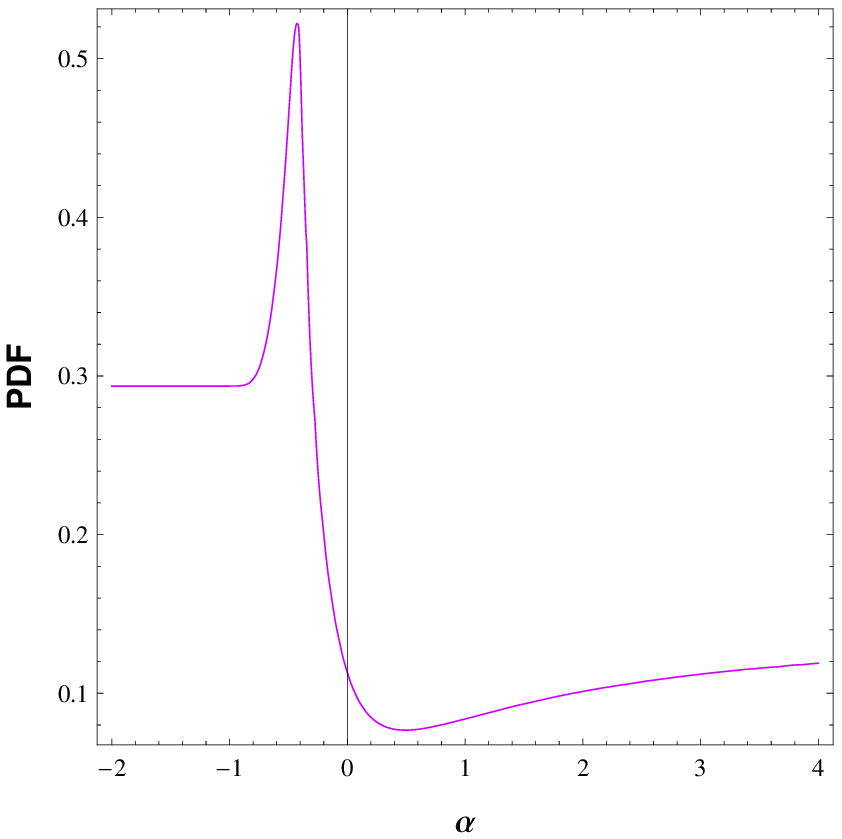}
\end{minipage} \hfill
\caption{{\protect\footnotesize One dimensional PDFs for the parameter $\alpha$ in the Rastall's scalar model with $\bar A = 0.1$; $\bar A = 0.5$ and $\bar A = 0.9$.
}}
\label{rastall-fig2}
\end{figure}
\end{center}
\begin{center}
\begin{figure}[!t]
\begin{minipage}[t]{0.3\linewidth}
\includegraphics[width=\linewidth]{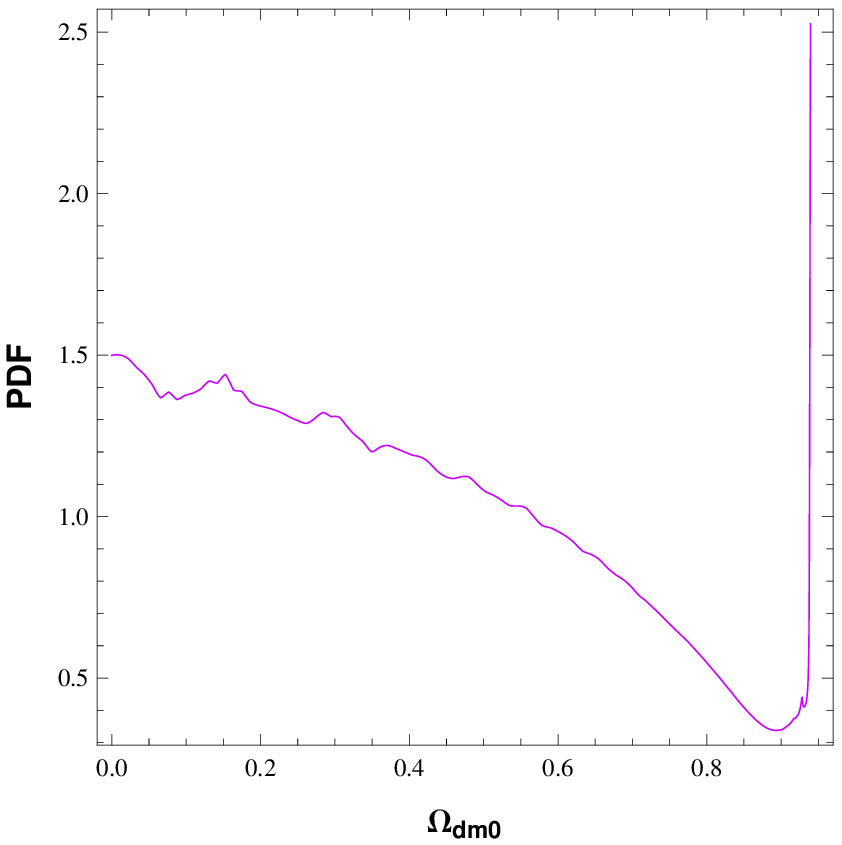}
\end{minipage} \hfill
\begin{minipage}[t]{0.3\linewidth}
\includegraphics[width=\linewidth]{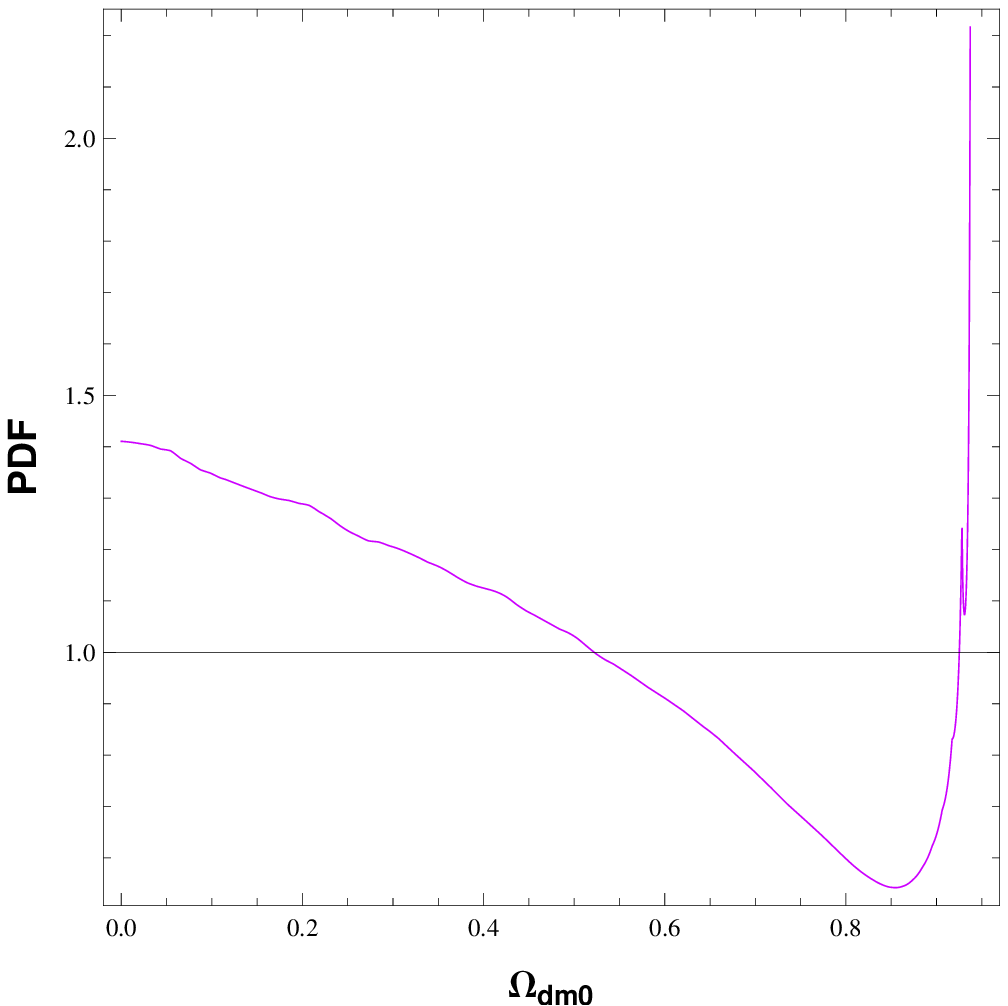}
\end{minipage} \hfill
\begin{minipage}[t]{0.3\linewidth}
\includegraphics[width=\linewidth]{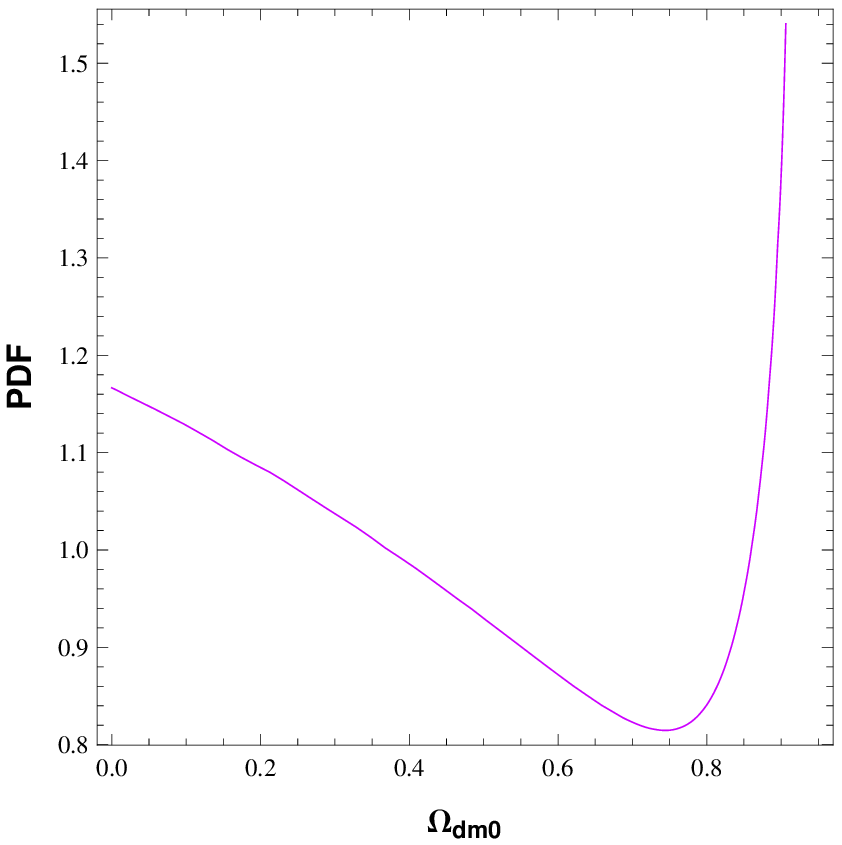}
\end{minipage} \hfill
\caption{{\protect\footnotesize One dimensional PDFs for the parameter $\Omega_{dm0}$ in the Rastall's scalar model with $\bar A = 0.1$; $\bar A = 0.5$ and $\bar A = 0.9$.
}}
\label{rastall-fig3}
\end{figure}
\end{center}

The model is tested against the 2dFGRS power spectrum data. We implement the initial condition as described in the previous section.
The same considerations can be made, grosso modo, about the behavior of the potential of the scalar field. The statistical analysis is
made in the same lines as before. First, we leave two free parameters, $\alpha$ and $\Omega_{dm0}$ and fix $\bar A$.
When $\bar A = 0.1, 0.5$ and $0.9$ the results are displayed in figures (\ref{rastall-fig1}, \ref{rastall-fig2},\ref{rastall-fig3}). For $\alpha$ the results are similar to those found previously with
the canonical scalar field. But, for $\Omega_{dm0}$, there are now two maximum of probabilities, one when the dark matter component is absent, and the other
when dark matter is the only component besides baryons. The second maximum is higher than the first one. In all these cases, the $AIC$ parameter is
about $20$. If, as before, $\alpha$ is restricted to be less than $- 1$, the results are essentially the same, including for the $AIC$ parameter.
\par
If the unification scenario is imposed from the beginning, leaving $\bar A$ and $\alpha$ as the free parameters, the behavior for $\alpha$ remains the same, while the most probable value for $\bar A$ is near $\bar A = 0.004$ - this leads to a an almost $CDM$ scenario, with a cosmological term very
near zero. This result can modified as the numerical precision is increased but $\bar A = 0$ (exactly $CDM$ model) seems not to
be the favored case. Again, the $AIC$ parameter is about $20$.
\par
With respect to the $\Lambda CDM$ model, the difference in the $AIC$ parameter is $\Delta AIC \sim 3$. Following the Jeffrey's scale this give
a moderate support in favor or the $\Lambda CDM$ compared with the Rastall's scalar model for the generalized Chaplygin gas \cite{liddlebis,marek}. Perhaps, due to details concerning the implementation of
this scale \cite{liddlebis,marek}, it would be more appropriate to state that both models remain competitive as far as the structure formation test is concerned.

\begin{center}
\begin{figure}[!t]
\begin{minipage}[t]{0.3\linewidth}
\includegraphics[width=\linewidth]{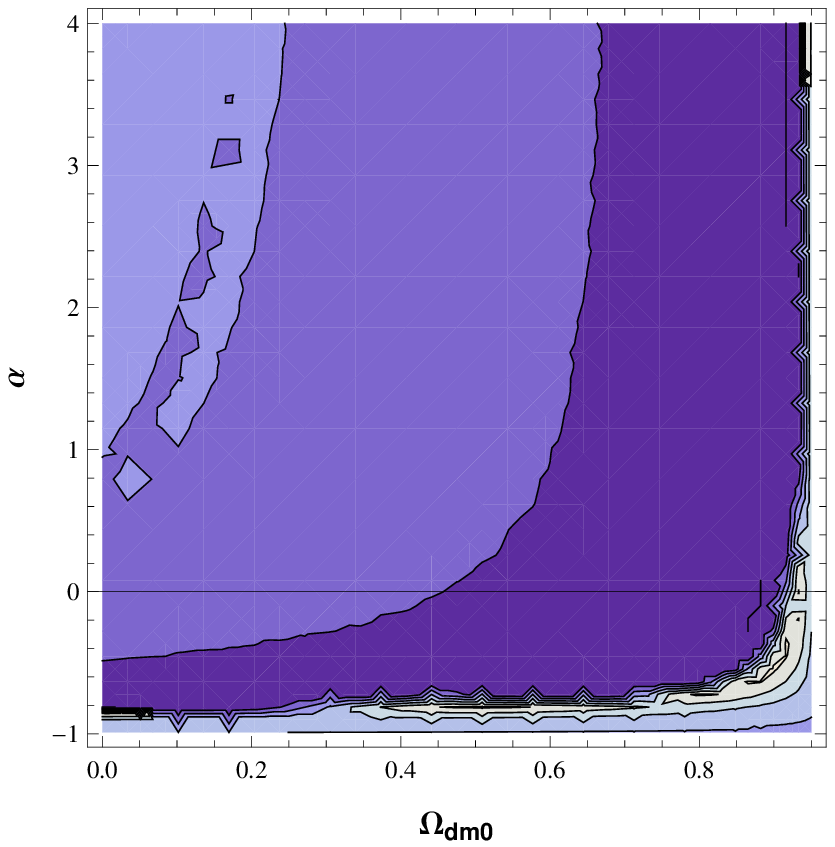}
\end{minipage} \hfill
\begin{minipage}[t]{0.3\linewidth}
\includegraphics[width=\linewidth]{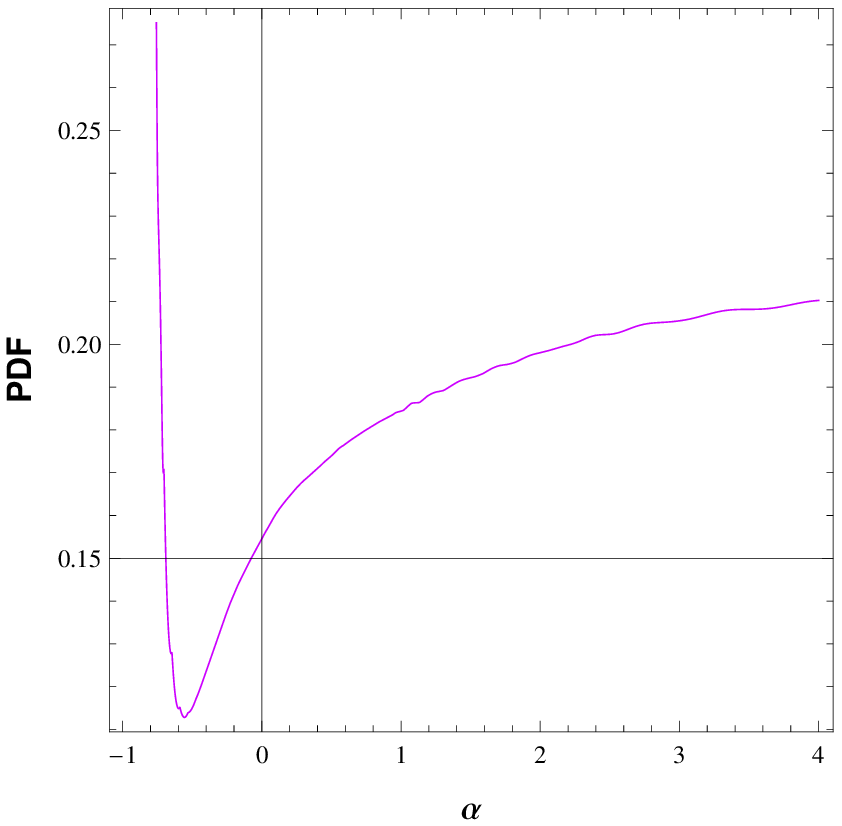}
\end{minipage} \hfill
\begin{minipage}[t]{0.3\linewidth}
\includegraphics[width=\linewidth]{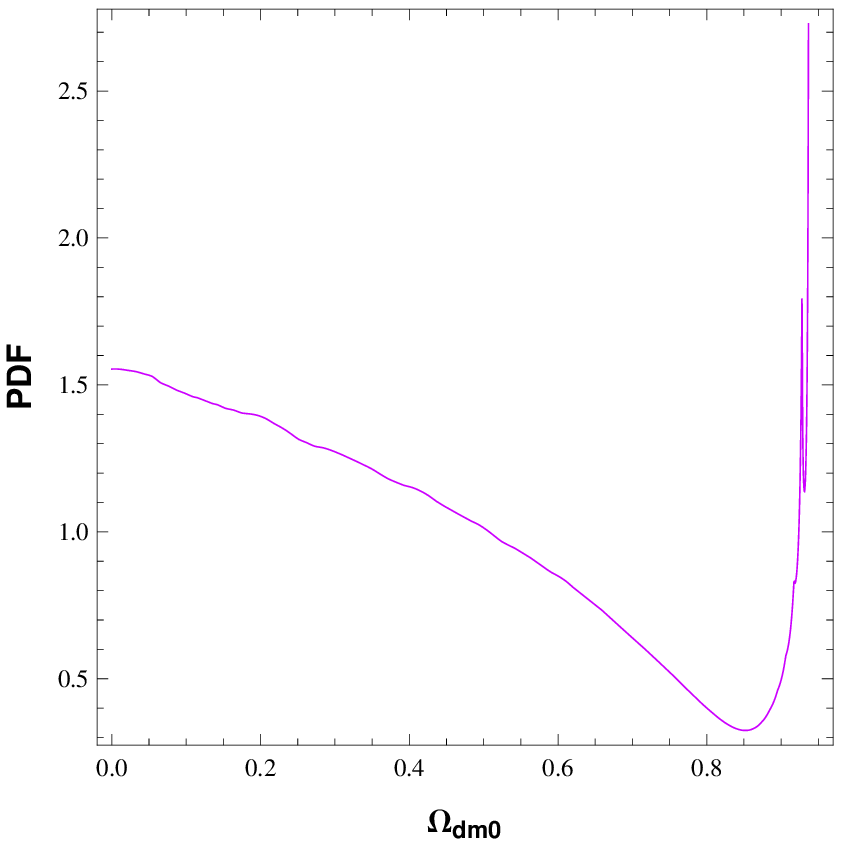}
\end{minipage} \hfill
\caption{{\protect\footnotesize Two and one dimensional PDFs restricting $\alpha > - 1$ and fixing $\bar A = 0.5$ for the Rastall's scalar field model.
}}
\label{-1-rastall}
\end{figure}
\end{center}
 
\begin{center}
\begin{figure}[!t]
\begin{minipage}[t]{0.3\linewidth}
\includegraphics[width=\linewidth]{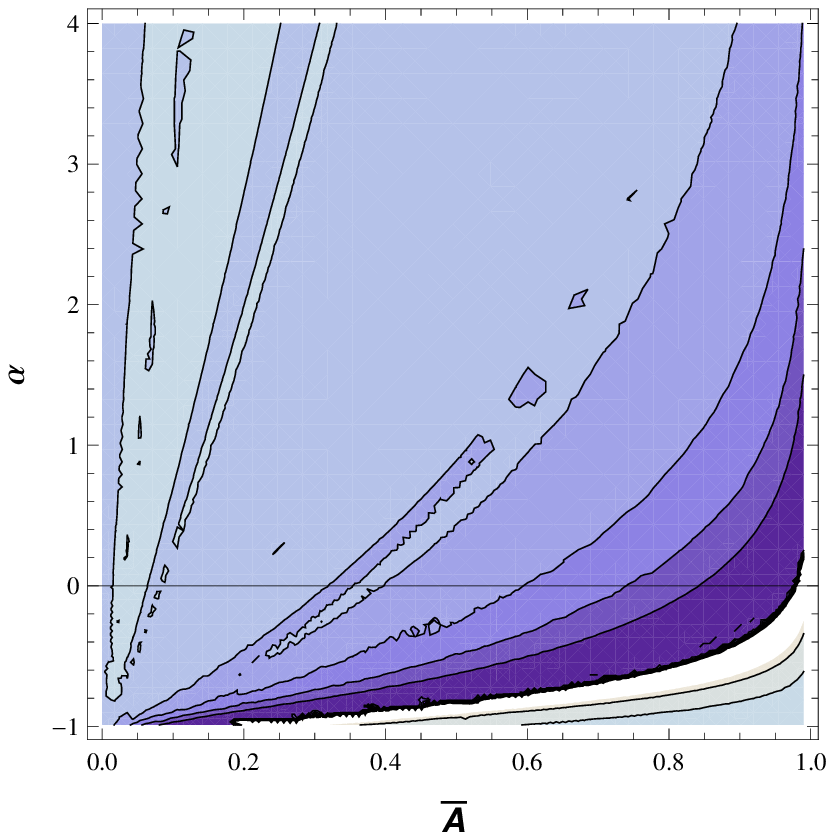}
\end{minipage} \hfill
\begin{minipage}[t]{0.3\linewidth}
\includegraphics[width=\linewidth]{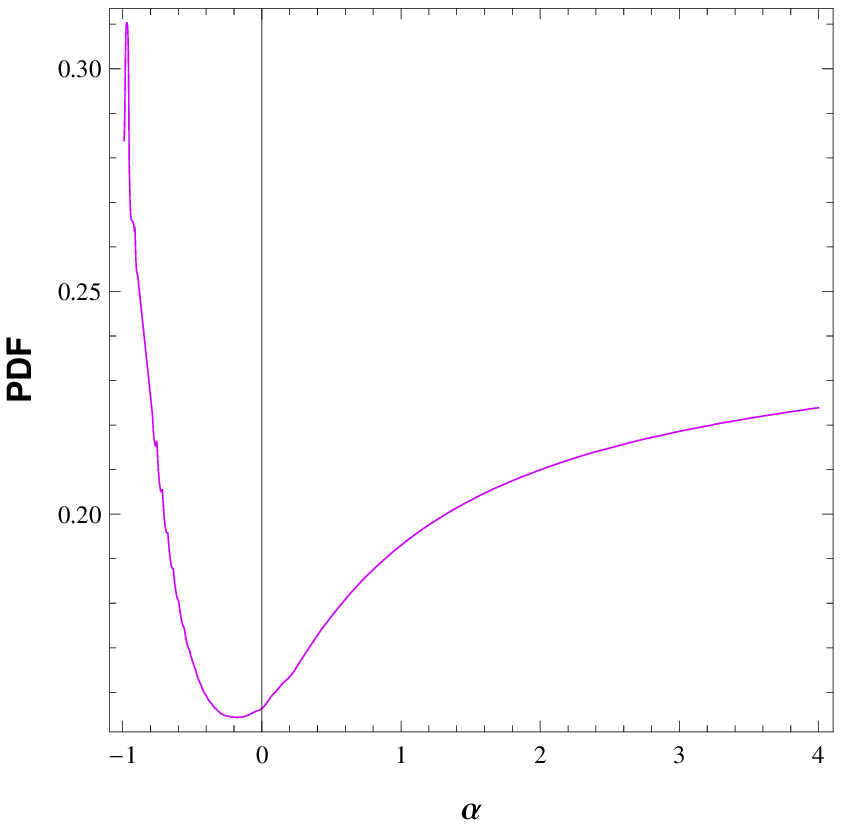}
\end{minipage} \hfill
\begin{minipage}[t]{0.3\linewidth}
\includegraphics[width=\linewidth]{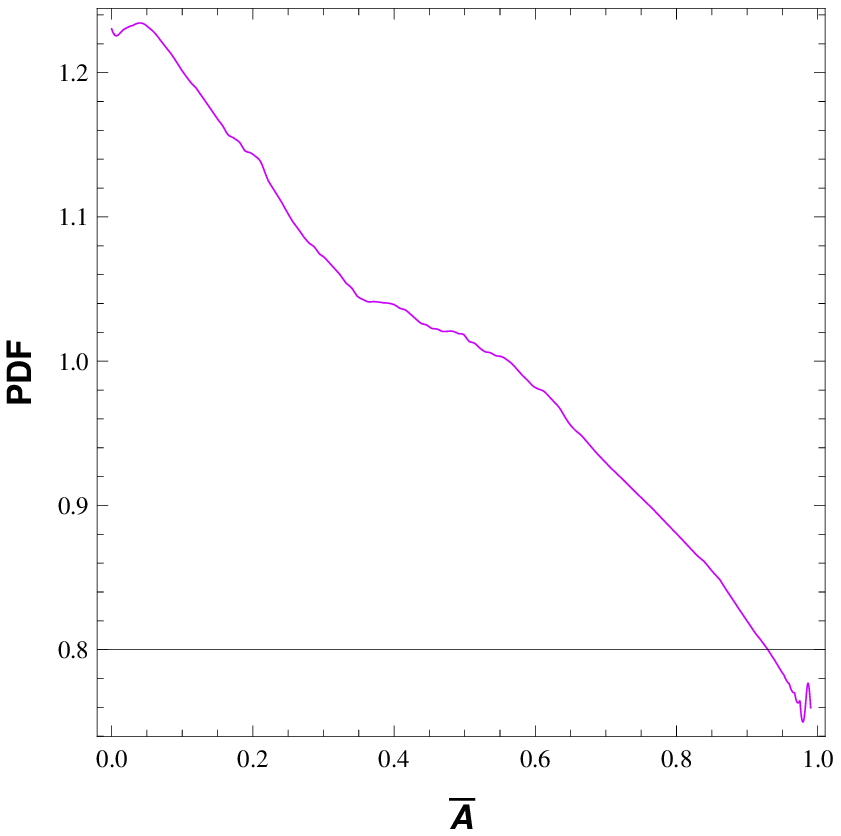}
\end{minipage} \hfill
\caption{{\protect\footnotesize Two and one dimensional PDFs restricting $\Omega_{dm0} = 0$ for the Rastall scalar field model.
}}
\label{rastall-unif}
\end{figure}
\end{center}

\section{Conclusions}

We have investigated here the possibility that the Generalized Chaplygin Gas (GCG) model may be represented by a self-interacting scalar field, instead
of using the fluid representation or the Dirac-Born-Infeld (DBI) action for a tachyonic fluid. The main reason for this investigation is
that both the fluid or DBI representations forbid to extend the analysis of the CGG model to negative values of $\alpha$ at perturbative level, due
to a negative squared sound velocity, which drives strong instabilities. This represents an important restriction, mainly when it is taken into
account that some background observational tests, like SN Ia, favor negative values of $\alpha$. 
\par
We have shown initially that, for the GCG model, the fluid representation is equivalent to the tachyonic representation of the
generalized DBI action,
at background and perturbative levels, extending the results of reference \cite{eduardo}. Hence, in order to consider the possibility of negative values for $\alpha$ in a perturbative analysis, we must use another framework, which we have chosen, first, that one represented 
by a canonical self-interacting scalar field, and later by a non-canonical scalar field inspired in the Rastall's theory of gravity. In these cases, negative values for $\alpha$ do not lead to any problem connected with the
sign of the (squared) sound velocity. When baryons are presented, there is no closed form for the potential representing the GCG in both scalar
models, but
an implicit expression, using the scale factor as variable, can be obtained.
\par
We have made a statistical analysis using the perturbative expressions for the self-interacting scalar field model for the GCG, computing the
matter power spectrum for the matter component, and confronting it with the 2dFGRS observational data. In principle, we allow the presence
of dark matter besides baryons and the scalar field. For a canonical scalar field, the constraints obtained indicate that a kind of quintessence scenario emerges,
with $\Omega_{dm0} \sim 0.23$ (like in the $\Lambda$CDM model). High positive values of $\alpha$ are favored. The best fitting configuration
implies typically $\chi^2_{min} \sim 0.30$, better than in the case of the $\Lambda$CDM model, for which $\chi^2_{min} \sim 0.38$.
An interesting scenario emerges when the unification scenario is imposed from the beginning, fixing $\Omega_{dm0} = 0$. The best
fitting is $\chi_{min}^2 \stackrel{>}{\sim} 0.30$, still smaller than in the $\Lambda$CDM case, but now for $\alpha$ negative. 
\par 
The above described scalar model has an important drawback: the velocity of light is equal to one, and it cannot represent dark matter at
the end of the radiative era, and it must be seen as an effective model valid during the matter dominated era, asking for a coherent description for
the previous era. Hence, another model was introduced based on the scalar field version of the Rastall's theory. For this case, the velocity
of the sound can be set equal to zero, independently of the potential term chosen. For this non-canonical scalar field, the quintessence model
is not anymore favored and, when the dark matter density parameter is free, there are two maximums, one for a universe entirely filled by
dark matter (in agreement with the power spectrum results for the GCG in the DBI representation), and the other when dark matter is absent
(in agreement with SN Ia results). The best fit is achieved, generally, with $\chi_{min}^2 \stackrel{>}{\sim} 0.38$, similar to the
$\Lambda CDM$ case, but with one more free parameter. Due to this, and employing the AIC criteria and the Jeffrey's scale to evaluate the
evidence in favor of one model or other, supports marginally the $\Lambda CDM$ model, but without a decisive result: the scalar field model remains
competitive. All these results are valid even when the unification dark matter/dark energy prior is imposed from the beginning.
\par
For the parameter $\alpha$ there is a qualitatively agreement with the SN Ia analysis.
Perhaps this may indicate a new concordance model, different from the $\Lambda$CDM. However, a more deep perturbative
analysis must be performed, mainly using the Integrated Sachs-Wolfe effect (a very delicate test for the unified models \cite{bertacca,barrow}), or even the full anisotropy spectrum of CMB. We hope to present
this analysis in the future.

\section*{Acknowledgements}

We thank CNPq (Brazil) for partial financial support.


%

\begin{thebibliography}{99}
\bibitem{komatsu} E. Komatsu et al., {\it Seven-year Wilkinson microwave anisotropy probe (WMAP) observations: cosmological interpretation}, arXiV:1001.4538.
\bibitem{caldwell} R.R. Caldwell and M. Kamionkowski, Ann. Rev. Nucl. Part. Sci. {\bf 59}, 397(2009).
\bibitem{bertone}  G. Bertone, D. Hooper and J. Silk, Phys. Rep. {\bf 405}, 279(2005). 
\bibitem{padma}  T. Padmanabhan, Phys. Rep. {\bf 380}, 235(2003). 
\bibitem{martin} J. Martin, Mod. Phys. Lett. {\bf A23}, 1252(2008).
\bibitem{moschella} A.Y. Kamenshchik, U. Moschella and V. Pasquier, Phys. Lett.
{\bf B511}, 265(2001).
\bibitem{berto1}
M.C. Bento, O. Bertolami and A.A. Sen, Phys. Rev. {\bf D66},
043507 (2002).
\bibitem{neven}  N. Bilic, G.B. Tupper and R.D. Viollier, Phys. Lett. {\bf B535}, 17(2002). 
\bibitem{fabris} J.C. Fabris, S.V.B. Gon\c{c}alves and P.E. de Souza, Gen. Rel. Grav. {\bf 34}, 53(2002). 
\bibitem{jackiw} R. Jackiw, {\it A particle field theorist's lectures on supersymmetric, non abelian fluid mechanics and d-branes},
physics/0010042. 
\bibitem{colistete} R. Colistete Jr, J. C. Fabris, S.V.B. Gon\c{c}alves and P.E. de Souza, Int. J. Mod. Phys. D13,
669(2004); R. Colistete Jr., J. C. Fabris and S.V.B. Gon\c{c}alves, Int. J. Mod. Phys. D14,
775(2005); R. Colistete Jr. and J. C. Fabris, Class. Quant. Grav. 22, 2813(2005).
\bibitem{berto2}  T. Barreiro, O. Bertolami and P. Torres, Phys. Rev. {\bf D78}, 043530(2008).
\bibitem{finelli}  L. Amendola, F. Finelli, C. Burigana and D. Carturan, JCAP {\bf 0307}, 005(2003).
\bibitem{piattella1} O. Piattella, JCAP {\bf 1003}, 012(2010). 
\bibitem{piattella2}  V. Gorini, A. Y. Kamenshchik, U. Moschella, O. F. Piattella and A. A. Starobinsky, JCAP {\bf 0802}, 016(2008). 
\bibitem{hermano} J.C. Fabris, S.V.B. Gon\c{c}alves, H.E.S. Velten and W. Zimdahl, Phys. Rev. D78, 103523
(2008). J.C. Fabris, H.E.S. Velten and W. Zimdahl, Phys. Rev. D81, 087303(2010).
\bibitem{eduardo} C.E.M. Batista, J.C. Fabris and M. Morita, Gen. Rel. Grav. {\bf 42}, 839(2010).
\bibitem{rastall} P. Rastall, Phys. Rev. {\bf D6}, 3357 (1972).
\bibitem{liddle}  C. Gao, M. Kunz, A.R. Liddle and D. Parkinson, Phys. Rev. {\bf D81}, 043520(2010).
\bibitem{gorini} V. Gorini, A.Y. Kamenshchik, U. Moschella and V. Pasquier, Phys. Rev. {\bf D69}, 123512(2004).
\bibitem{frolov} A.V. Frolov, L. Kofman and A.A. Starobinsky, Phys. Lett. {\bf B545}, 8(2002).
\bibitem{gori} V. Gorini, A. Kamenshchik, U. Moschella, V. Pasquier and A. Starobinsky, Phys. Rev. {\bf D72}, 103518(2005).
\bibitem{bbks} N. Sugiyama, Astrophys. J.Suppl. {\bf 100}, 281(1995); J.M. Bardeen, J.R. Bond, N. Kaiser and A.S. Szalay, Astrophys. J. {\bf 304}, 15 (1986).
\bibitem{sola} J.C. Fabris, J. Sol\`a and I.L. Shapiro, JCAP {\bf 0702},016(2007). 
\bibitem{staro} A. Shafieloo, V. Sahni and A.A. Starobinsky, Phys. Rev. {\bf D80}, 101301(2009).
\bibitem{smalley} L.L. Smalley, Il Nuovo Cim. {\bf B80}, 42(1984).
\bibitem{hamani} C.E.M. Batista, J.C. Fabris e  M. Hamani Daouda, Il Nuovo Cim. {\bf B125}, 957(2010)
\bibitem{galileon}  A. Nicolis, R. Rattazzi and E. Trincherini, Phys. Rev. {\bf D79}, 064036(2009).
\bibitem{liddlebis} A.R. Liddle, Month. Not. R. Astron. Soc. {\bf 377}, L74(2007).
\bibitem{marek} M. Szydlowski and A. Kurek, {\it AIC, BIC, Bayesian evidence and a notion on simplicity of cosmological model}, arXiv:0801.0638.
\bibitem{bertacca} D. Bertacca and N. Bartolo, JCAP{\bf 0711}, 026(2007).
\bibitem{barrow}  B. Li and J.D. Barrow, Phys. Rev.{\bf D79},103521(2009).


\end{thebibliography}
\end{document}